\documentclass{JHEP3}
\usepackage{axodraw}
\keywords{QCD, Jets, Parton Model, Phenomenological Models}
\preprint{LU-TP 05-10\\
  hep-ph/0503181
}

\usepackage{cite}
\usepackage{epsfig}
\usepackage{xspace}
\usepackage{graphics}
\usepackage[latin1]{inputenc}

\def\hide#1{}

\newcommand{\ariadne}{A\scalebox{0.8}{RIADNE}\xspace}

\newcommand{\ab}{\ensuremath{\bar{\alpha}}}
\newcommand{\as}{\ensuremath{\alpha_{\mathrm{s}}}}

\newcommand{\kT}{\ensuremath{k_{\perp}}}
\newcommand{\pT}{\ensuremath{p_{\perp}}}

\newcommand{\kTi}[1]{\ensuremath{k_{\perp #1}}}
\newcommand{\pTi}[1]{\ensuremath{p_{\perp #1}}}

\newcommand{\particle}[1]{\ensuremath{\mathrm{#1}}}
\newcommand{\antiparticle}[1]{\ensuremath{\bar{\mathrm{#1}}}}
\newcommand{\el}{\particle{e}}

\newcommand{\g}{\particle{g}}

\newcommand{\q}{\particle{q}}

\newcommand{\qbar}{\antiparticle{q}}

\newcommand{\ee}{\ensuremath{\el^+\el^-}}
\newcommand{\tee}{\ensuremath{\el^+\el^-}\xspace}

\def\mrm#1{\mathrm{#1}}

\def\sub#1{\ensuremath{_{\mrm{#1}}}}

\def\f2d3{\ensuremath{F_2^{\mrm{D}3}}}
\def\ftwo{\ensuremath{F_2}\xspace}
\def\done#1{}

\newcommand{\eqref}[1]{eq.~(\ref{#1})\xspace}

\newcounter{aenumct}

{\end{list}}

\newcounter{enumct}

\skip\footins = 1\bigskipamount plus 2pt minus 4pt                              

\title{Energy Conservation and Saturation in Small-{\boldmath$x$} Evolution}

\author{Emil Avsar, Gösta Gustafson and Leif Lönnblad\\
  Dept.~of Theoretical Physics,
  Sölvegatan 14A, S-223 62  Lund, Sweden\\
  E-mail: \email{Emil.Avsar@thep.lu.se}, \email{Gosta.Gustafson@thep.lu.se}
    and \email{Leif.Lonnblad@thep.lu.se}}
  
  \abstract{Important corrections to BFKL evolution are obtained from
    non-leading contributions and from non-linear effects due to
    unitarisation or saturation. It has been difficult to estimate the
    relative importance of these effects, as NLO effects are most
    easily accounted for in momentum space while unitarisation and
    saturation are easier in transverse coordinate space. An essential
    component of the NLO contributions is due to energy conservation
    effects, and in this paper we present a model for implementing
    such effects together with saturation in Mueller's dipole
    evolution formalism.  We find that energy conservation severely
    dampens the small-$x$ rise of the gluon density and, as a
    consequence, the onset of saturation is delayed. Using a simple
    model for the proton we obtain a reasonable qualitative
    description of the $x$-dependence of \ftwo at low $Q^2$ as
    measured at HERA even without saturation effects.  We also give
    qualitative descriptions of the energy dependence of the cross
    section for $\gamma^\star\gamma^\star$ and $\gamma^\star-$nucleus
    scattering.  }

\begin{document}
 
\sloppy
 
\section{Introduction}
\label{sect-intro}

In the asymptotical high-energy limit, QCD should be described by
BFKL\cite{Kuraev:1977fs,Balitsky:1978ic} evolution, at least to
leading order and assuming a fixed coupling. Here terms of the form
$(\as\log x)^n$ in a perturbative expansion are resummed to all
orders. The result is a fast rise of any cross section with increasing
energy or, equivalently, with decreasing $x$. The rise has the form
$x^{-\lambda}$, where the power $\lambda$ to leading order is given by
$\ab \,4\log2$, which is around one half for $\ab\equiv3\as/\pi=0.2$.
Clearly such a behavior will violate the unitarity bound for large
enough energies. To cure this problem Gribov, Levin and Ryskin
\cite{Gribov:1984tu} in pioneering works discussed non-linear effects
from gluon recombination, which cause the gluon density to saturate
before it becomes too high.

Because the transverse coordinates are unchanged in a high energy
collision, unitarity constraints are generally more easy to take into
account in a formalism based on the transverse coordinate space
representation, and several suggestions for how to include saturation
effects in such a formalism have been proposed. Golec--Biernat and
W\"usthoff \cite{Golec-Biernat:1998js} formulated a dipole model, in
which a virtual photon is treated as a $q\bar{q}$ or $q\bar{q}g$
system impinging on a proton, and this approach has been further
developed by several authors (see e.g.\ \cite{Forshaw:1999uf} and
\cite{Bartels:2002cj}).  Mueller \cite{Mueller:1993rr,Mueller:1994jq,Mueller:1994gb}
has formulated a dipole cascade model in transverse coordinate space,
which reproduces the BFKL equation, and in which it is also possible
to account for multiple sub-collisions.  Within this formalism
Balitsky and Kovchegov \cite{Balitsky:1995ub,Kovchegov:1999yj} have
derived a non-linear evolution equation, which also takes into account
these saturation effects from multi-pomeron exchange. In an
alternative approach a high density gluonic system is described by a
so-called Color Glass Condensate \cite{Iancu:2000hn,Ferreiro:2001qy},
where non-perturbative effects appear due to the high density, even
though the coupling \as\ is small.

There are, however, other effects which may dampen the growth of
the structure function. One is the fact that the next-to-leading
logarithmic corrections to the BFKL evolution turn out to be very
large\cite{Fadin:1998py,Ciafaloni:1998gs}. These corrections strongly
suppress the growth for small $x$, and in fact, even for moderate
values of \ab, the power $\lambda$ becomes negative.
It is well-known \cite{Salam:1999cn} that a major fraction of these
higher order corrections is related to energy conservation. The large
effect of energy-momentum conservation is also clearly demonstrated by
the numerical analyses by Andersen--Stirling \cite{Andersen:2003gs} and
Orr--Stirling \cite{Orr:1997im}.

As a consequence there is currently some controversy over whether
saturation has been observed in deeply inelastic scattering at HERA.
Saturation effects have been studied in the coordinate space
representation in which it has been difficult to include non-leading
effects, and the non-leading effects have been studied in momentum
space, where it is hard to include saturation. Therefore it has been
difficult to estimate the relative importance of saturation and
non-leading effects. To know if the dominant mechanism behind the
reduced growth rate is due to energy conservation or to saturation is
also very important for reliable extrapolations to higher energies at
LHC and high energy cosmic ray events. Our aim in this paper is to
find a formalism where it is possible to account for both
unitarisation and energy-momentum conservation, knowing that the
latter is a major part of the non-leading effects. An alternative
approach to this problem is presented in
\cite{Levin:1999mw,Gotsman:2005vc}, in which a formalism to include
saturation and conservation of energy (or rather the positive
lightcone momentum component, $p_+$) is described. In our formalism we
emphasize conservation of both lightcone components, $p_+$ and $p_-$.

We emphasize that the question concerning saturation is not whether it
exists in general --- clearly if the scale is small enough so that the
transverse size of the gluons is as big as a nucleon there must be
recombinations present --- rather the debate is about whether effects
of recombination of \textit{perturbative} gluons at scales above a
couple of GeV has been observed. The deviation from the linear BFKL
evolution, as a consequence of saturation, is expected to be essential
below a line $Q^2=Q_s^2(x)$ in a ($Q^2,x$)
diagram\cite{Golec-Biernat:1998js}. The effect can be viewed in two
different ways, as a suppression of the logarithmic $x$-slope of the
structure function, $d\log\ftwo/d\log x \equiv
\lambda_{\mathrm{eff}}$, when $x$ becomes small for fixed $Q^2$, or as
a suppression when $Q^2$ becomes small for fixed $x$. HERA data show
an almost linear dependence of $\lambda_{\mathrm{eff}}$ with $\log
Q^2$, leveling off at $\approx0.1$ for $Q^2$ below 1~$\mathrm{GeV}^2$,
with the proviso that the $x$-interval used to determine the slope is
not constant, but is shifted towards smaller $x$ for smaller
$Q^2$-values (see e.g.\ refs.~\cite{Adloff:2001rw,DIS04Petrukhin}).
The suppression for small $x$ and for small $Q^2$ also appears to
satisfy a scaling property called geometric scaling, saying that \ftwo
is a function of a single variable $\tau=Q^2/Q_s^2(x)$. This scaling
is satisfied by the HERA data to a high degree, and in an early study
Golec--Biernat and W\"usthoff found a good fit to experiments with
$Q_s^2(x)=(3.04\cdot10^{-4}/x)^{0.288}\, \mathrm{GeV}^2$
\cite{Golec-Biernat:1998js}.  In a more recent analysis Iancu,
Itakura, and Munier \cite{Iancu:2003ge} obtained a good fit to later
HERA data with a model based on BFKL evolution including some
non-leading effects\footnote{Basically, non-leading effects are taken
  into account by simply lowering the BFKL $\lambda$, or treating it
  as a free parameter, in which case it comes out close to the value
  predicted by the so-called renormalization-group
  improved\cite{Ciafaloni:1999yw} NLO BFKL. Also some non-leading
  effects are introduced by letting \as\ run, typically with $Q^2_s$
  as the scale.}  plus saturation, with
$Q_s^2(x)=(0.267\cdot10^{-4}/x)^{0.253}\, \mathrm{GeV}^2$. This value
is smaller than the one in ref.~\cite{Golec-Biernat:1998js}, and
therefore moves the saturation region closer to the non-perturbative
regime.

The Mueller dipole evolution is formulated in rapidity
($\propto\log1/x$) and transverse coordinate space, with rapidity
acting as the evolution parameter.  A DIS $\gamma^\star p$ scattering
is typically viewed in the rest system of the proton, where the
$\gamma^\star$ evolves into a $\q\qbar$ pair, long before the
interaction. This $\q\qbar$ pair will then radiate off gluons,
$\q\qbar\to\q\g\qbar\to\q\g\g\qbar\to\ldots$, a process which is
formulated in terms of radiation from colour-dipoles. The initial
dipole between the \q\ and \qbar\ emits a gluon, splitting the dipole
into two, one between the \q\ and \g\ and one between the \g\ and
\qbar, both of which can continue radiating gluons. In the end, one of
these dipoles can interact with the proton, giving a cross section
which increases
as a power of $1/x$, and in leading order reproduces the BFKL result.
Saturation comes in because when the density of dipoles becomes large
there is a possibility that more than one of them interacts with the
proton, thus slowing down the increase of the cross section. This
effect can be interpreted as multi-pomeron exchange, and is taken into
account in the non-linear BK equation.

The Mueller dipole evolution is very similar in spirit to the Dipole
Cascade Model (DCM)\cite{Gustafson:1986db,Gustafson:1988rq}, which
describes time-like evolution of dipoles from e.g.\ an initial
$\q\qbar$ pair created in \tee-annihilation. However, here the
evolution is formulated in momentum space. The transverse momentum is
used as evolution parameter, and the conservation of energy and
momentum is simple to implement, especially in a Monte Carlo Event
Generator. This model gives a very good description of \tee
annihilation and the \ariadne program\cite{Lonnblad:1992tz}, where it
is implemented, describes almost all data from the four LEP
collaborations to an astonishing precision (see e.g.\ 
\cite{Hamacher:1995df}). Also with the so-called soft-radiation
extension of the DCM, DIS final states as measured by HERA are well
described using a simple semi-classical description of time-like
dipole evolution (see e.g.\ \cite{Brook:1995nn}).

One problem in Mueller's formulation is the fact that, while the
emission probability for a time-like cascade in the DCM is finite for a
fixed value of the evolution parameter, the emission probability here
diverges $\sim 1/r^2$ for very small dipole sizes $r$. However, the
interactions from the colour charge and anti-charge interfere
destructively, resulting in colour transparency, and for small
$r$-values the dipole cross section is proportional to $r^2$, implying
that the total cross section remains finite, and the Mueller dipole
formulation can be shown to be equivalent to BFKL.  Although
$\sigma_\mathrm{tot}$ is finite, the singularities do cause problems.
For a numerical analysis or a MC simulation it is necessary to
introduce a cutoff for small dipoles, and for small cutoff values the
number of dipoles becomes very large. In fact, the increase is so
strong that a Monte Carlo simulation of the evolution, as is done
e.g.\ in the OEDIPUS program \cite{Salam:1996nb,Mueller:1996te,Salam:1995uy},
becomes extremely inefficient. It also implies that it is not possible
to interpret the dipole chain as a real final state.  If a small size
in coordinate space corresponds to a large transverse momentum, the
very large and diverging number of dipoles with very small sizes
obviously violates energy-momentum conservation. Instead these
emissions have to be regarded as virtual fluctuations, which in
Mueller's approach are handled by appropriate Sudakov form factors.

An alternative approach to DIS is the so-called Linked Dipole Chain
(LDC) \cite{Andersson:1996ju,Andersson:1998bx} model, where an initial
set of gluons is obtained using space-like parton evolution, and then
is evolved in time-like cascades into final-state gluons. LDC is a
reformulation and generalization of CCFM evolution
\cite{Catani:1990yc,Ciafaloni:1988ur}, which reproduces BFKL in the
asymptotic small-$x$ limit but is also similar to DGLAP evolution
\cite{Altarelli:1977zs,Gribov:1972ri,Lipatov:1975qm,Dokshitzer:1977sg}
at larger $x$ values. In addition to sequences of DGLAP evolution,
where the \textit{upward} gluon branchings with $\kTi{i}\gg\kTi{i-1}$
are strongly ordered in rapidity and in the \kT\ of the propagating
gluon, also \textit{downward} splittings with $\kTi{i}\ll\kTi{i-1}$
may appear with a reduced weight. The result is an evolution which is
totally symmetric, in the sense that it can be generated either from
the projectile or from the target end with the same result. The LDC
model is completely formulated in momentum space which makes it easy
to implement in a Monte Carlo event generator
\cite{Kharraziha:1998dn}, where energy and momentum conservation is
easily accomplished.

In this paper we will identify some similarities between the LDC model
and the Mueller dipoles, and use them to derive a scheme for
implementing energy momentum conservation in the space-like dipole
evolution. We conjecture that only gluon emissions which satisfy
energy-momentum conservation can correspond to real final state
gluons, and that keeping only these (with a corresponding
modification of the Sudakov form factor) will not only give a better
description of the final states, but also account for essential parts
of the NLO corrections to the BFKL equation. Our approach is based on
the observation that the emission of a dipole with a very small
transverse size, $r$, corresponds to having two very well localized
gluons, and such gluons must have large transverse momenta, of the
order of $\pT\sim1/r$. By in this way assigning a transverse momentum
to each emitted gluon, and also taking into account the recoils of the
emitting gluons, we can then make sure that each dipole splitting is
kinematically allowed. However, as will be discussed in detail in
section~\ref{sec:final-states}, energy-momentum conservation is a
\textit{necessary} condition for a chain to correspond to a real final
state, but it is not a \textit{sufficient} condition. Therefore we
will in this paper only discuss results for total cross sections, and
postpone discussions of final state properties to a future
publication.

The program described here is, of course, not easy to implement in an
analytic formalism.  Instead we have written a Monte Carlo program,
similar to OEDIPUS, where the kinematics can be easily treated.  This
program can then be used to calculate cross sections for e.g.\ 
dipole--dipole scattering at different virtualities. We also introduce
a simple model for nucleons as a distribution in dipole numbers and
sizes, to investigate cross sections for dipole--$A$ scattering.
In principle this can also be used to study $AA$ scattering, but such
investigations will also be postponed for a future publication.

The layout of this paper is as follows. First we describe the dipole
cascades formulated both in transverse momentum and in coordinate
space in sections~\ref{sec:dipole-models-momentum-space} and
\ref{sec:dipole-models-coordinate-space}. In
section~\ref{sec:energy-momentum-conservation} we then describe the
similarity between the LDC model and Mueller's cascade model, and how
this guides us in the introduction of energy-momentum conservation
into the Mueller dipole formalism.  In this section we also discuss
some open questions related to final state properties and gluon
recombination. In section~\ref{sec:monte-carlo-impl} we describe
briefly the implementation in a Monte Carlo program we use to obtain
the results presented in the following section~\ref{sec:results}.
Finally we arrive at our conclusions in section \ref{sec:conclusions}.

\done{Still expect large NLO effects. These can be tamed by including
  energy conservation.}

\done{We have built a dipole MC, cf.\ Salam, with energy conservation
  inspired by LDC is built in.}

\done{Layout of paper.}

\section{Dipole Cascades in Momentum Space}
\label{sec:dipole-models-momentum-space}

The Dipole Cascade Model (DCM\cite{Gustafson:1986db,Gustafson:1988rq})
as implemented in the \ariadne\cite{Lonnblad:1992tz} event generator
has been very successful in describing the bulk of the data on
hadronic final states recorded at LEP. In this section we will first
describe briefly this model and then go on to how it can be extended
to also describe cross sections and hadronic final states in DIS by a
reformulation of the CCFM evolution.

\subsection{Time-like Cascades}
\label{sec:timelike-cascades}

In \tee annihilation, the emission of a gluon from the initial
\q\qbar-pair can be described in terms of dipole radiation from the
colour-dipole between the \q\ and \qbar. Subsequent emission of a
second gluon is then described as radiation from either of the two
dipoles connecting the quark with the gluon and the gluon with the
anti-quark.  In the dipole rest frame the relative probability for
such a dipole splitting is to leading logarithmic order given by the
standard dipole radiation formula
\begin{equation}
  \label{eq:diprad}
  d{\cal P}\propto\as\frac{d\kT^2}{\kT^2}dy.
\end{equation}
The available phase space is a triangular region in the ($\log\kT^2,y$)
plane, $\kT e^{\pm y}<W$, where $W$ is the invariant mass of the
dipole.

Clearly this is very similar to the Mueller dipole formalism. The main
differences are that here we have dipoles in momentum space rather
than in transverse position, and the evolution is in decreasing
transverse momentum rather than in increasing rapidity.  Hence we here
have a Sudakov form factor
\begin{equation}
  \label{eq:dipsud}
  -\log\Delta_S(\kTi{\max}^2,\kT^2)=
  \int_{\kT^2}^{\kTi{\max}^2}\frac{d{\cal P}}{d\kT^{'2}}d\kT^{'2}.
\end{equation}
Also, we here deal only with real final-state emissions, while Mueller's
formalism describe initial-state virtual dipoles.

The ordering in decreasing \kT\ (measured in the rest frame of the
emitting dipole) means that energy and momentum conservation is a
relatively small correction.  This formalism is easily implemented in
a Monte Carlo generator, in which it is straight forward to take into
account non-leading corrections to the emission probability in
eq.~(\ref{eq:diprad}) and energy-momentum conservation including proper
recoils of the emitters, which modifies the triangular phase space
region slightly as shown in figure \ref{fig:phasespace}.

This formalism has many advantages as compared to conventional parton
cascades. One very essential feature is that coherence effects,
conventionally implemented as angular ordering, is automatically taken
into account in a way which is more accurate than the conventional
sharp angular cut.

\FIGURE{
  \epsfig{file=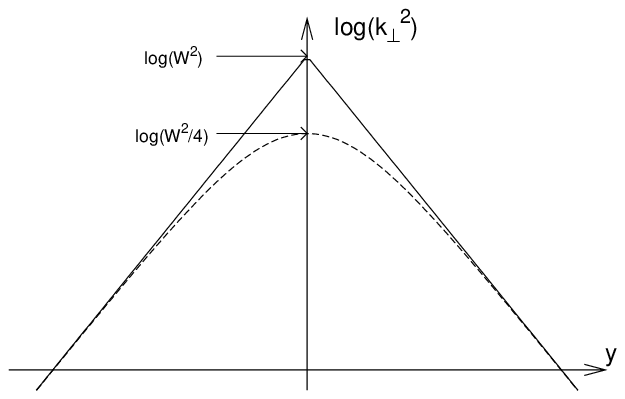,width=10cm}
  \caption{\label{fig:phasespace} The available phase space for a
    gluon emitted with some transverse momentum \kT\ and rapidity $y$
    from a dipole with total invariant mass $W$. The full line
    represents the approximate phase space limits relevant for a
    leading log calculation, while the dashed line represents the
    modification when taking recoils of the emitting quarks into
    account.}}

\done{DCM $\ee\to\q\qbar\to\q\g\qbar\to\q\g\g\qbar\to\ldots$}

\subsection{Space-like Cascades}
\label{sec:spacelike-cascades}

The dipole cascade model has been extended to also describe deeply
inelastic lepton--hadron collisions in two different ways. The one
which is implemented in \ariadne relies on a semi-classical model
\cite{Andersson:1989gp} where all gluon emissions are treated as
final-state radiation. This has been very successful in describing
hadronic final states at HERA, but suffers from the fact that it does
not predict the cross section. It is also difficult to relate to any
standard evolution equation, although it has qualitative similarities
with BFKL and CCFM.

The other extension is called the Linked Dipole Chain (LDC) model
\cite{Andersson:1996ju} and uses a reformulation and generalization of
CCFM evolution to build up an initial set of gluon emissions, which
determines the cross section.  These gluons define a chain of
linked dipoles, which may initiate standard final-state dipole
splittings, which then do not affect the cross section. The initial
gluons are carefully selected to be purely real final-state gluons,
i.e.\ only such emissions are considered which are not accompanied by
large virtual corrections given by the so-called non-Sudakov form
factor in CCFM (or Regge form factor in BFKL).

It turns out that these emissions are those where the gluons are
ordered in both positive and negative light-cone momenta and with
transverse momenta which are larger than the smaller of transverse
momenta of the radiating propagator gluon before and after the
emission: $\pTi{i}>\min(\kTi{i-1},\kTi{i})$. We are then left with
simple splittings which either increase the \kT\ of the propagator,
given by
\begin{equation}
  \label{eq:splitup}
  d{\cal P}=\ab\frac{d\kTi{i}^2}{\kTi{i}^2}\frac{dz_i}{z_i},
\end{equation}
or decreasing it, given by
\begin{equation}
  \label{eq:splitdown}
  d{\cal P}=\ab\frac{d\kTi{i}^2}{\kTi{i}^2}\frac{dz_i}{z_i}
  \frac{\kTi{i}^2}{\kTi{i-1}^2}.
\end{equation}
The extra suppression $\kTi{i}^2/\kTi{i-1}^2$ for evolution with
decreasing \kT\ ensures that the evolution becomes symmetric, i.e.\ it
does not matter whether we evolve from the proton or the virtual
photon end. A local maximum, $k_{\perp \mathrm{max}}$, can be interpreted 
as evolutions from the projectile and target ends up to a central hard 
sub-collision. If treated as evolution from one end we then have a step up to
$k_{\perp \mathrm{max}}$ followed by a step down in $k_\perp$, and from
the weights in eqs.~(\ref{eq:splitup}) and (\ref{eq:splitdown}) this
gives the correct factor $1/k_{\perp \mathrm{max}}^4$ expected from 
Rutherford scattering. If we instead have a local minimum, 
$k_{\perp \mathrm{min}}$, then there is no associated power of
$k_\perp$, and such a minimum is therefore free of singularities.

Also the LDC model has been implemented in a Monte Carlo generator
\cite{Kharraziha:1998dn}, which fairly well reproduces final states at
HERA. However, there is a caveat, namely that crucial measurements
sensitive to small-$x$ dynamics, such as the rates of forward jets,
can only be reproduced if non-singular parts of the gluon splitting
function are omitted. For further discussions on this subject, we
refer the reader to ref.\ \cite{Andersson:2002cf}.

\done{LDC  $\gamma^\star\to\q\qbar\to\q\g\qbar\to\q\g\g\qbar\to\ldots$.}



\section{Dipole Cascades in Coordinate Space}
\label{sec:dipole-models-coordinate-space}

\subsection{The Mueller Dipole Formulation}
\label{sec:muell-dipole-form}

\FIGURE{
\scalebox{0.8}{\mbox{
\begin{picture}(400,100)(0,0)
\Vertex(50,100){2}
\Vertex(50,0){2}
\Vertex(150,100){2}
\Vertex(150,0){2}
\Vertex(180,30){2}
\Vertex(285,100){2}
\Vertex(285,0){2}
\Vertex(315,30){2}
\Vertex(315,90){2}
\Line(50,0)(50,100)
\DashLine(150,100)(150,0){2}
\Line(150,100)(180,30)
\Line(150,0)(180,30)
\DashLine(285,100)(285,0){2}
\DashLine(285,100)(315,30){2}
\Line(285,100)(315,90)
\Line(315,90)(315,30)
\Line(285,0)(315,30)
\Text(40,100)[]{$Q$}
\Text(40,0)[]{$\bar{Q}$}
\Text(60,100)[]{$1$}
\Text(60,0)[]{$0$}
\Text(140,100)[]{$1$}
\Text(140,0)[]{$0$}
\Text(140,50)[]{$\mathbf{r}_{01}$}
\Text(190,30)[]{$2$}
\Text(175,78)[]{$\mathbf{r}_{12}$}
\Text(175,2)[]{$\mathbf{r}_{02}$}
\Text(275,100)[]{$1$}
\Text(275,0)[]{$0$}
\Text(325,30)[]{$2$}
\Text(325,90)[]{$3$}
\LongArrow(75,50)(110,50)
\LongArrow(210,50)(245,50)
\LongArrow(360,25)(360,50)
\LongArrow(360,25)(385,25)
\Text(360,60)[]{$y$}
\Text(395,25)[]{$x$}
\end{picture}
}}
\caption{\label{figdipolesplit} A quark-antiquark dipole in transverse
  coordinate space is split into successively more dipoles via gluon
  emission.}}
Consider now the process of an evolving onium state or $\gamma^*
\rightarrow Q\bar{Q} \rightarrow Q g \bar{Q} \rightarrow Q g g \bar{Q}
\rightarrow \ldots$ in transverse coordinate space, as illustrated in
figure~\ref{figdipolesplit}.  Here a virtual photon is split into a
$Q\bar{Q}$ colour dipole, which is first split into two dipoles by the
emission of a gluon, then into three dipoles by a second gluon, etc.
The probability for such a dipole splitting is given by the expression
(for notation see figure~\ref{figdipolesplit})
\begin{eqnarray}
\frac{dP}{dy} &=& \frac{\bar{\alpha}}{2\pi} d^2 \textbf{r}_2
\frac{r_{01}^2}{r_{02}^2 \,r_{12}^2} \cdot S \nonumber \\
\mathrm{where} \,\,\,
S &=& \exp\left[-\frac{\bar{\alpha}}{2\pi} \int dy\int d^2\textbf{r}_2
\frac{r_{01}^2}{r_{02}^2 \,r_{12}^2}\right].
\label{splitprob}
\end{eqnarray}
Here $S$ denotes a Sudakov form factor. When this dipole splitting is
iterated it evolves into a cascade with with an exponentially
increasing number of dipoles.

We note that the weight in eq.~(\ref{splitprob}) is singular, and the
integral over $d^2\textbf{r}_2$ in the Sudakov form factor diverges
for small values of $r_{02}$ and $r_{12}$. Therefore Mueller
introduced a cutoff $\rho$, such that the splitting has to satisfy
$r_{02}>\rho$ and $r_{12}>\rho$. The integral in $S$ is then also
restricted in the same way.  A small cutoff value $\rho$ will here
imply that we get very many dipoles with small $r$-values. However, as
the cross section for a small dipole interacting with a target also
gets small (see below), the total cross section is finite also in the
limit $\rho \rightarrow 0$.

A proton target can be treated as a collection of dipoles. When
two dipoles collide, there is a recoupling of the colour charges,
forming new dipole chains. This is schematically illustrated in
figure \ref{figgammagamma} for the case of $\gamma^\star \gamma^\star$
scattering. Here we imagine the two virtual photons splitting up into
quark-antiquark pairs, which develop into two colliding dipole
cascades. When the two central dipoles collide, it implies a
recoupling, as indicated by the arrow.  The weight for this
interaction is given by the expression \cite{Salam:1995uy}
\begin{equation}
  f = \frac{\alpha_s^2}{2} \left\{\log\left[
      \frac{|\textbf{r}_1 -\textbf{r}_3|\cdot|\textbf{r}_2 -\textbf{r}_4|}
      {|\textbf{r}_1 -\textbf{r}_4|\cdot|\textbf{r}_2 -\textbf{r}_3|}
    \right]\right\}^2.
\label{dipoltvarsnitt}
\end{equation}
An important property of this expression is that when e.g.\ the left
of the interacting dipoles is small, the weight in
eq.~(\ref{dipoltvarsnitt}) can be shown to be proportional to
$(\mathbf{r}_1 - \mathbf{r}_2)^2$, which compensates the factor
$(\mathbf{r}_1 - \mathbf{r}_2)^{-2}$ in the evolution probability from
eq.~(\ref{splitprob}).

\FIGURE{
\scalebox{0.75}{\mbox{
\begin{picture}(400,175)(0,5)
\Photon(20,130)(60,130){2}{4}
\Photon(380,140)(340,140){2}{4}
\DashLine(60,130)(90,145){2}
\DashLine(60,130)(90,115){2}
\DashLine(340,140)(310,160){2}
\DashLine(340,140)(310,120){2}
\DashLine(90,145)(90,115){2}
\DashLine(310,120)(310,160){2}
\DashLine(90,115)(120,140){2}
\DashLine(120,140)(145,118){2}
\DashLine(160,124)(120,140){2}
\DashLine(310,120)(290,156){2}
\DashLine(310,120)(260,163){2}
\DashLine(260,163)(250,124){2}
\DashLine(250,124)(235,160){2}
\DashLine(235,160)(225,120){2}
\ArrowLine(90,145)(120,140)
\ArrowLine(120,140)(170,165)
\ArrowLine(145,118)(90,115)
\ArrowLine(170,165)(160,124)
\ArrowLine(160,124)(145,118)
\ArrowLine(120,140)(170,165)
\ArrowLine(290,156)(310,160)
\ArrowLine(260,163)(290,156)
\ArrowLine(310,120)(250,124)
\ArrowLine(225,120)(220,161)
\ArrowLine(250,124)(225,120)
\ArrowLine(220,161)(235,160)
\ArrowLine(235,160)(260,163)
\Vertex(310,160){2}
\Vertex(310,120){2}
\Vertex(90,145){2}
\Vertex(90,115){2}
\Vertex(120,140){2}
\Vertex(170,165){2}
\Vertex(145,118){2}
\Vertex(160,124){2}
\Vertex(290,156){2}
\Vertex(260,163){2}
\Vertex(310,120){2}
\Vertex(250,124){2}
\Vertex(235,160){2}
\Vertex(260,163){2}
\Vertex(220,161){2}
\Vertex(225,120){2}
\Text(10,130)[]{$\gamma^*$}
\Text(390,140)[]{$\gamma^*$}
\Text(90,155)[]{$Q$}
\Text(90,105)[]{$\bar{Q}$}
\Text(310,170)[]{$\bar{Q}'$}
\Text(310,110)[]{$Q'$}
\Text(210,161)[]{$r_3$}
\Text(215,120)[]{$r_4$}
\Text(180,165)[]{$r_1$}
\Text(170,124)[]{$r_2$}
\LongArrow(190,105)(190,75)
\DashLine(170,65)(160,24){2}
\ArrowLine(160,24)(145,18)
\ArrowLine(120,40)(170,65)
\DashLine(225,20)(220,61){2}
\ArrowLine(250,24)(225,20)
\ArrowLine(220,61)(235,60)
\Vertex(220,61){2}
\Vertex(225,20){2}
\Vertex(170,65){2}
\Vertex(160,24){2}
\ArrowLine(170,65)(220,61)
\ArrowLine(225,20)(160,24)
\Text(170,75)[]{$r_1$}
\Text(160,14)[]{$r_2$}
\Text(220,71)[]{$r_3$}
\Text(225,10)[]{$r_4$}
\LongArrow(300,50)(300,75)
\LongArrow(300,50)(325,50)
\Text(300,85)[]{$\mathbf{r}$}
\Text(350,43)[]{$y=$rapidity}
\end{picture}
}}
\caption{\label{figgammagamma}A symbolic picture of a $\gamma^* \gamma^*$
  collision in rapidity-$\mathbf{r}_\perp$-space. The two dipole
  chains interact and recouple with probability $f$ given by
  eq.~(\ref{dipoltvarsnitt}).  }}

A $\gamma^\star p$ collision is frequently analyzed in the rest frame
of the target proton. When the virtual photon hits the target, the
number of dipoles present in this frame grows in accordance with the
BFKL equation, and the total cross section increases proportional to
$\exp(\lambda Y)$, where the total rapidity interval $Y$ is given by
$Y = \log(1/x) = \log(s/Q^2)$.

It is, however, also possible to study the collision in a frame
different from the target rest frame. Then the target dipoles evolve
in the same way a distance $y$ in rapidity, while the projectile
dipole evolves the shorter distance $Y-y$. As the evolution grows
exponentially with rapidity, the cross section is proportional to
$\exp(\lambda y) \cdot \exp(\lambda(Y-y)) = \exp(\lambda Y)$, which
means that it is insensitive to the chosen frame, in which the
collision is studied.  This frame independence is, however, broken by
multiple collision effects related to unitarity. This will be
discussed further in section~\ref{sec:frame-dependence}.

\subsection{Unitarity and Saturation}
\label{unitarity-saturation}

\FIGURE{
  \epsfig{file=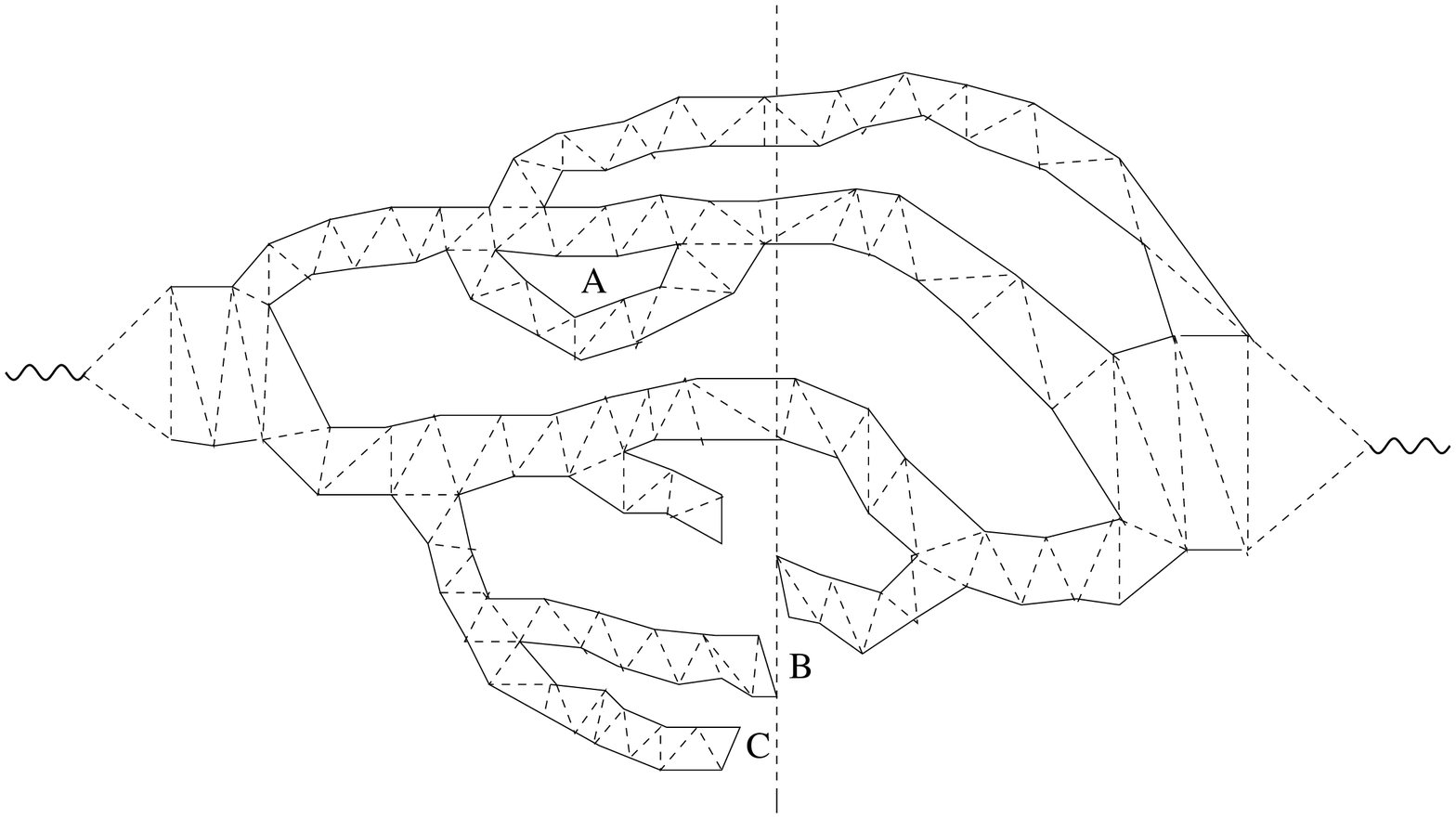,width=10cm}
  \caption{\label{fig:multcoll}A $\gamma^\star  \gamma^\star$ collision
    event with multiple sub-collisions in
    rapidity-$\mathbf{r}_\perp$-space. At high energies several
    branches from the two colliding dipole systems may reconnect. The
    dashed vertical line symbolizes the Lorentz frame in which the
    collision is evaluated.}
  \label{multcoll}}

A great advantage of the coordinate space representation is the fact
that the transverse coordinate $\mathbf{r}$ is unchanged during the
collision.  This implies that unitarity can very easily be implemented
by the replacement $f \rightarrow 1-e^{-f}$.  As the dipole cascades
from the two virtual photons branch out, it is also possible to have
{\em multiple interactions} with dipoles from the left and from the
right, as illustrated in figure \ref{fig:multcoll}. The total cross
section is then given by
\begin{equation}
\sigma \propto \int d^2\mathbf{b} (1-e^{-\sum f_{ij}}).
\label{multint}
\end{equation}
where $\mathbf{b}$ denotes the impact parameter separation between the
two initial particles, and the sum runs over all pairs $i$ and $j$ of
colliding dipoles. Here the factor $1-e^{-\sum f_{ij}}$, where the
exponent corresponds to a no-interaction probability, ensures that the
unitarity constraint is satisfied. The first term in an expansion,
$\sum f_{ij}$, corresponds to a single pomeron exchange, while the
higher order terms are related to multi-pomeron exchanges.

Including these non-linear terms in the evolution equation leads to
the Balitsky--Kovchegov (BK)
equation\cite{Balitsky:1995ub,Kovchegov:1999yj}. The BK equation
governs the small-$x$ evolution of the $F_2$ structure function of a
large nucleus. In his original paper Kovchegov assumed a target
nucleus at rest and an evolved projectile dipole. Using Mueller's
dipole formulation for the evolution of the dipole and summing pomeron
exchanges of all orders he derived the following equation:
\begin{equation}
  \frac{dN(\mathbf{r}_{01},Y)}{dY}=
  \frac{\bar{\alpha}}{2\pi}\int d^2\mathbf{r}_2 
  \frac{r_{01}^2}{r_{02}^2r_{12}^2}
  (N(\mathbf{r}_{12},Y)+N(\mathbf{r}_{02},Y)-N(\mathbf{r}_{01},Y)
  -N(\mathbf{r}_{12},Y)N(\mathbf{r}_{02},Y)).
\label{BK1}
\end{equation}
Here $N(\mathbf{r}_{ij},Y)$ denotes the forward scattering amplitude
(which also determines the total reaction probability) of the dipole
$\mathbf{r}_{ij}$ on the target nucleus. The nucleus has been assumed
to be large, which means that the impact parameter dependence of $N$
is suppressed.

For small $N$-values the quadratic term can be neglected, and
eq.~(\ref{BK1}) is reduced to Mueller's linear equation for the dipole
cascade evolution.  This equation is just the BFKL equation formulated
in the dipole language.  The first two terms correspond to the
emission of a gluon forming two new dipoles, while the term with a
minus sign accounts for the virtual corrections described by the
Sudakov form factor in eq.~(\ref{splitprob}). The quadratic term
represents the effect of multiple collisions, which become more
important when $N$ becomes large. This suppresses the growth rate for
larger $Y$-values and results in saturation when $N$ approaches 1,
thus preserving unitarity.

The BK equation (\ref{BK1}) describes the development of the cascade
before it hits a dense nuclear target. It can also be used to describe
the evolution of two colliding cascades in a
$\gamma^\star\gamma^\star$ collision, as illustrated in figure
\ref{fig:multcoll}.  Here several branches from the two systems may
reconnect as described in figure \ref{figgammagamma} and
eq.~(\ref{dipoltvarsnitt}).  We note here that the cascade evolution
described by the linear terms in eq.~(\ref{BK1}) are only leading in
colour, while the effect from multiple collisions is formally colour
suppressed. Therefore this formalism includes corrections from
multiple sub-collisions in the Lorentz frame in which the process is
evaluated (denoted by the vertical dashed line in figure
\ref{fig:multcoll}), but does not take into account the possibility
that two branches recombine before the collision. Such an event is
indicated by the letter $A$ in figure \ref{fig:multcoll}. This effect
is also colour suppressed and thus not included in the evolution. As a
consequence the result depends on the Lorentz frame used, and this
problem will be further discussed in
section~\ref{sec:frame-dependence}.







\section{Combining Energy-Momentum Conservation and Unitarity}
\label{sec:energy-momentum-conservation}

With a small cutoff $\rho$ ($r>\rho$) we get, as mentioned above, very
many small dipoles. If these are interpreted as real emissions, with
transverse momenta proportional to $1/r$, it would imply a
catastrophic violation of energy-momentum conservation. As discussed
above, the emission of these small dipoles have a very limited effect
on the total cross section, and they have to be interpreted as virtual
fluctuations.  Thus the result in eq.~(\ref{multint}) will describe
the inclusive cross section, but the many dipoles produced in all the
branching chains will not correspond to the production of exclusive
final states.

\subsection{Relation Mueller's Dipole Cascade vs. LDC}
\label{mueller-ldc}

Before a discussion of these virtual fluctuations we want to discuss
the relation between Mueller's Dipole Cascade and the LDC model. Let
us study the chain of emissions, which is illustrated in
figure \ref{figdglap}.  Apart from the Sudakov factors this chain gets
the following weight:
\begin{eqnarray}
\frac{d^2 \textbf{r}_2\, r_{01}^2}{r_{02}^2 \,r_{12}^2} \cdot 
\frac{d^2 \textbf{r}_3\, r_{12}^2}{r_{13}^2\, r_{23}^2} 
\cdot \frac{d^2 \textbf{r}_4\, r_{23}^2}{r_{24}^2 \,r_{34}^2} \cdot 
\frac{d^2 \textbf{r}_5\, r_{34}^2}{r_{35}^2 \,r_{45}^2} \cdot
\frac{d^2 \textbf{r}_6\, r_{35}^2}{r_{36}^2 \,r_{56}^2}= \nonumber \\
= r_{01}^2 \frac{d^2 \textbf{r}_2 \,
d^2 \textbf{r}_3 \,d^2 \textbf{r}_4\, d^2 \textbf{r}_5\, d^2 \textbf{r}_6}
{r_{02}^2 \,r_{13}^2\,r_{24}^2\,r_{45}^2\,r_{36}^2 \,r_{56}^2}
\label{chainweight}
\end{eqnarray}
We here note that the total weight is a product of factors
$1/r_{ij}^2$ for all \textit{``remaining dipoles''}, i.e.\ for those
dipoles which have not been split by further gluon emission. They are
marked by solid lines in figure \ref{figdglap}. All dependence on the
size of \textit{``intermediate''} dipoles, which have disappeared
because they split in two daughter dipoles, is canceled in
eq.~(\ref{chainweight}), as they appear both in the numerator and in
the denominator. (These dipoles are marked by dashed lines in
figure \ref{figdglap}.)

If a dipole size, $\textbf{r}$, is small, it means that the gluons are
well localized, which must imply that transverse momenta are
correspondingly large.  This implies that not only the new gluon gets
a large $k_\perp \sim 1/r$, but also that the original gluon, which is close in
coordinate space, gets a corresponding recoil. For the special example
in figure \ref{figdglap} the emissions of the gluons marked 2, 3, and 4
give dipole sizes which become smaller and smaller, $a>>b>>c>>d$, in
each step of the evolution.  (This also implies that the ``remaining''
and the ``intermediate'' dipoles are pairwise equally large.) The
corresponding $k_\perp$-values therefore become larger and larger in
each step. After the minimum dipole, with size $d$, the subsequent
emissions, 5, and 6, give again larger dipoles with correspondingly
lower $k_\perp$ values.  The probability for this chain is
proportional to
\begin{equation}
\frac{d^2 \textbf{r}_2}{b^2} \cdot 
\frac{d^2 \textbf{r}_3}{c^2} 
\cdot \frac{d^2 \textbf{r}_4}{d^0} \cdot 
\frac{d^2 \textbf{r}_5}{e^2} \cdot
\frac{d^2 \textbf{r}_6}{f^2}\cdot\frac{1}{f^2} 
\label{dglap}
\end{equation}

For the first emissions, 2 and 3, we in this expression recognize the
product of factors $\prod d^2\textbf{r}_i/r_i^2\, \propto \prod
d^2\textbf{k}_i/k_i^2$, just as is expected from a ``DGLAP evolution''
of a chain with monotonically increasing $k_\perp$. Emission number 4
corresponds to the minimum dipole size, $d$, which should be associated with
a maximum $k_\perp$. In the following evolution the dipole sizes get 
larger again, corresponding to successively smaller transverse momenta.
In analogy with the evolution in the LDC model described in 
section~\ref{sec:spacelike-cascades}, this latter
part can be interpreted as DGLAP evolution in the opposite direction,
i.e. from the target end up to the central hard sub-collision. 
In this sub-collision the gluons 3 and 4 recoil 
against each other with transverse momenta $k_{\perp \mathrm{max}}$. 
In eq.~(\ref{dglap}) we see that the
factors of $d$ have canceled, which thus gives
the weight $d^2 \textbf{r}_4 \propto
d^2\textbf{k}_{\mathrm{max}}/k_{\mathrm{max}}^4$. This reproduces the weight 
expected from a hard gluon--gluon scattering, and corresponds
exactly to the result in the LDC model as discussed in 
section~\ref{sec:spacelike-cascades}.

\FIGURE{
\scalebox{0.9}{\mbox{
\begin{picture}(400,83)(0,13)
\Vertex(20,50){2.5}
\Vertex(250,70){2.5}
\Vertex(240,30){1}
\Vertex(260,35){1}
\Vertex(260,25){1}
\Vertex(290,25){1}
\Vertex(335,70){1}
\DashLine(20,50)(250,70){6}
\Line(20,50)(240,30)
\DashLine(250,70)(240,30){2}
\Line(250,70)(260,35)
\DashLine(240,30)(260,35){2}
\DashLine(260,35)(260,25){2}
\Line(240,30)(260,25)
\Line(260,25)(290,25)
\DashLine(260,35)(290,25){2}
\Line(290,25)(335,70)
\Line(260,35)(335,70)
\Text(135,70)[]{$a$}
\Text(135,30)[]{$\approx a$}
\Text(10,50)[]{$0$}
\Text(235,50)[]{$\approx b$}
\Text(263,56)[]{$b$}
\Text(250,78)[]{$1$}
\Text(236,24)[]{$2$}
\Text(264,41)[]{$3$}
\Text(260,18)[]{$4$}
\Text(293,19)[]{$5$}
\Text(335,78)[]{$6$}
\Text(250,36)[]{$c$}
\Text(250,22)[]{$c$}
\Text(265,30)[]{$d$}
\Text(279,35)[]{$e$}
\Text(276,20)[]{$e$}
\Text(316,44)[]{$f$}
\end{picture}
}}
%
\caption{\label{figdglap}A dipole cascade in
  rapidity-$\mathbf{r}_\perp$-space, where a chain of smaller and
  smaller dipoles is followed by a set of dipoles with increasing
  sizes. The initial dipole between points 0 and 1 is marked by long
  dashes, and those dipoles which have split into two new dipoles and
  disappeared from the chain are marked by short dashes. This chain is
  interpreted as one $k_\perp$-ordered cascade from one side and one
  from the other, evolving up to a central hard sub-collision, which
  is represented by the dipole with minimum size and therefore maximum
  $k_\perp$.}}

\FIGURE{
\begin{picture}(300,80)(0,0)
\DashLine(20,30)(20,50){4}
\Line(20,50)(80,60)
\DashLine(20,30)(80,60){2}
\Line(80,60)(220,70)
\DashLine(20,30)(220,70){2}
\DashLine(220,70)(260,52){2}
\Line(220,70)(268,60)
\Line(20,30)(260,52)
\Line(260,52)(268,60)
\Vertex(20,30){2}
\Vertex(20,50){2}
\Vertex(80,60){1}
\Vertex(260,52){1}
\Vertex(220,70){1}
\Vertex(268,60){1}
\Text(10,30)[]{$0$}
\Text(10,50)[]{$1$}
\Text(80,67)[]{$2$}
\Text(262,45)[]{$4$}
\Text(273,62)[]{$5$}
\Text(220,77)[]{$3$}
\Text(50,60)[]{$a$}
\Text(60,45)[]{$a$}
\Text(150,70)[]{$b$}
\Text(130,57)[]{$b$}
\Text(150,35)[]{$b$}
\Text(250,68)[]{$c$}
\Text(240,56)[]{$c$}
\Text(268,55)[]{$d$}
\end{picture}
\caption{\label{figminkt}A cascade where the dipoles increase to a
  maximum, and then decrease.  Here the size of the largest dipole,
  denoted $b$, corresponds to the minimum $k_\perp$ in the chain.}}

Figure \ref{figminkt} shows instead a chain with increasing dipole sizes
up to a maximum value, $r_{\mathrm{max}}$, which thus corresponds to a
minimum transverse momentum, $k_{\perp \mathrm{min}}$. Here we get the
weight $d^2 \textbf{r}_{\mathrm{max}}/r_{\mathrm{max}}^4 \propto d^2
\textbf{k}_{\mathrm{min}}$. Therefore there is no singularity for the
minimum $k_\perp$-value. This result is also directly analogous to
the corresponding result in the LDC model.

\subsection{Energy-Momentum Conservation}
\label{e-m-conservation}

As discussed in section \ref{sec:spacelike-cascades}, the main feature
of the LDC model is the observation that both the total cross section
and the final state structures are determined by chains consisting of
a subset of the gluons appearing in the final state. These gluons were
called ``primary gluons'' in ref.~\cite{Andersson:1996ju} and later
called ``backbone gluons'' in ref.~\cite{Salam:1999ft}. Remaining real
final state gluons can be treated as final state radiation from the
primary gluons. Such final state emissions do not modify the total
cross sections, and give only small recoils to the parent emitters.
The primary gluons have to satisfy energy-momentum conservation, and
are ordered in both positive and negative light-cone momentum
components, $p_+$ and $p_-$. We saw in the previous section that in
Mueller's cascade the emission probabilities for gluons, which satisfy
the conditions for primary gluons in LDC, have exactly the same
weight, when the transverse momenta are identified with the inverse
dipole size, $2/r$. This inspires the conjecture that with this
identification an appropriate subset of the emissions in Mueller's
cascade can correspond to the primary gluons in the momentum space
cascade, meaning that they determine the cross sections while the
other emissions can be regarded as either virtual fluctuations or
final state radiation.

A necessary condition for this subset of gluons is that energy and
momentum is conserved. (This is not a sufficient condition, as
discussed further below.) Only emissions which satisfy energy-momentum
conservation can correspond to real emissions, and keeping only these
emissions (with a corresponding modification of the Sudakov form
factor) gives a closer correspondence between the generated dipole
chains and the observable final states. To leading order this does not
change the total cross section. However, as it has been demonstrated
that a large fraction of the next to leading corrections to the BFKL
equation is related to energy conservation, we expect that taking this
into account will improve the results also in this dipole formulation
of the evolution.

A very important consequence of energy-momentum conservation is also
that it implies a {\em dynamical cutoff}, $\rho(\Delta y)$, which is
large for small steps in rapidity\footnote{Note that in our notation,
  $y$ is rapidity and not $\log(1/x)$.}, $\Delta y$, but gets smaller
for larger $\Delta y$.  (Alternatively it could be described as a
cutoff for $\Delta y$ which depends on $r$.) The production of a small
dipole with size $r$ corresponds to the emission of a gluon with
$k_\perp \approx 2/r$ and therefore $k_+ \approx (2/r) e^{-y}$.  Thus
conservation of positive light-cone momentum implies
\begin{equation}
r > 2e^{-\Delta y} / k_{\perp\mathrm{parent}}.
\end{equation}

Conserving also the negative light-cone momentum, $p_-$, implies that
we in a similar way also get a maximum value for $r$ in each emission.
Here we note that while the projectile has a large $p_+$ component and
a very small $p_-$ component, the target has small $p_+$ but
contributes (almost) all $p_-$. Thus conservation of $p_-$ means that
in the evolution of the projectile cascade, the $p_-$ components
become steadily larger, presuming that in end the collision with the
target will provide the total $p_-$ needed to put the cascade on
shell. (The kinematical details will be discussed further in
section~\ref{sec:monte-carlo-impl}.) Branches which do not interact
must consequently be regarded as virtual fluctuations, which are not
realized in the final state.

The net result of conservation of both $p_+$ and $p_-$ is that the
number of dipoles grows much more slowly with energy, and we will see
in section \ref{sec:results} that this also strongly reduces the total
cross sections. Besides this physical effect, it also simplifies the
implementation in a MC program, and implies that the severe numerical
complications encountered in MC simulations without energy
conservation, discussed in refs.~\cite{Salam:1995uy} and
\cite{Mueller:1996te}, can be avoided.

\subsection{Final States and Virtual Dipoles}
\label{sec:final-states}

However, even if we only include emissions which would be allowed by
energy-momentum conservation, this does not fully correspond to the
formation of a possible final state. As discussed above, the weight
contains in the denominator the square of all ``remaining dipoles''.
Even if the constraint from energy-momentum conservation implies a
minimum rapidity gap for the emission of small dipoles, this
suppression does not reproduce the weight $\propto d^2k_\perp /k_\perp^4$
for the smallest dipole in a sequence, needed to reproduce the cross
section for a hard sub-collision. A possible solution is to interprete
clusters of gluons, like those marked A, B, and C in figure
\ref{figeffectivegluons}, as ``effective gluons'', where the small
internal separations do not correspond to large transverse momenta for
real final state gluons. These hard emissions have to be compensated
by virtual corrections.  

From the weight in eq.~(\ref{splitprob}) we see that
that the emission probability, where such a small dipole is the
parent, is proportional to the square of its length, and therefore
suppressed.  However, if this dipole really does split by gluon
emission, and starts a branch which interacts and gets coupled to a
chain from the target (as illustrated in figure
\ref{figeffectivegluonsmult}) then the separation cannot be neglected.
In this case the two gluons at the dipole ends have to be treated as
independent, and can no longer be considered as a single effective
gluon. This problem concerning the properties of exclusive final
states will be further discussed in a forthcoming publication, and in
the following we will here only discuss results for total cross
sections.

A very important question concerns whether demanding energy conservation 
also for virtual emissions implies a serious overestimate of its 
consequences. The production of a small dipole implies large $p_\perp$ for 
the new gluon and also for its partner in the dipole, which suffers a recoil. 
Since the new dipole can be virtual only if it does not interact further,
this is a problem if the neighbouring dipoles are significantly changed 
by the emission. In 
the calculations of the effect of the recoil, the lightcone component 
$p_+ = e^{-y}\cdot p_\perp$ is conserved, which implies that the rapidity $y$ 
is adjusted to a larger value. This implies that the emission of the 
virtual dipole does not significantly modify the $p_+$-component of the 
neighbouring dipoles. It does, however, have a large effect on the 
values of $p_- = e^{+y}\cdot p_\perp$,
where the changes in $p_\perp$ and $y$ do not compensate each other. 
In order not to overestimate the effect of $p_-$-conservation, we therefore
in this analysis implement the constraint from $p_-$-conservation in such a 
way, that we neglect the size of the neighbouring dipoles. Thus we calculate 
this constraint assuming that the $p_\perp$ of the gluons in the dipole ends 
is determined only  by the size of the emitting dipole. This does 
somewhat underestimate the effect of $p_-$-conservation, but it avoids the 
large overestimate, which would be the consequence of including the 
unrealistic constraint from virtual dipole neighbours.

\FIGURE{
\scalebox{1.3}{\mbox{
\begin{picture}(200,100)(0,0)
\Vertex(20,20){2}
\Vertex(20,80){2}
\Vertex(40,75){1}
\Vertex(43,72){1}
\Vertex(45,25){1}
\Vertex(47,27){1}
\Vertex(49,28){1}
\Vertex(70,70){1}
\Vertex(72,68){1}
\Vertex(100,47){1}
\Vertex(120,60){1}
\Vertex(130,30){1}
\DashLine(20,20)(20,80){4}
\DashLine(20,20)(40,75){2}
\DashLine(20,20)(43,72){2}
\Line(20,20)(45,25)
\Line(20,80)(40,75)
\Line(40,75)(43,72)
\DashLine(43,72)(45,25){2}
\Line(47,27)(45,25)
\DashLine(43,72)(47,27){2}
\DashLine(43,72)(49,28){2}
\Line(47,27)(49,28)
\Line(43,72)(70,70)
\Line(70,70)(72,68)
\DashLine(49,28)(70,70){2}
\Line(49,28)(100,47)
\DashLine(49,28)(72,68){2}
\DashLine(100,47)(72,68){2}
\Line(120,60)(72,68)
\Line(120,60)(130,30)
\DashLine(120,60)(100,47){2}
\Line(100,47)(130,30)
\Text(42,82)[]{$A$}
\Text(49,20)[]{$B$}
\Text(70,77)[]{$C$}
\Text(15,20)[]{$0$}
\Text(15,80)[]{$1$}

\end{picture}
}}
\caption{\label{figeffectivegluons} The clusters of gluons marked
  A, B, and C must be interpreted as ``effective gluons''. The small
  dipole sizes do not correspond to large final state transverse
  momenta.}}

\FIGURE{
\scalebox{1.3}{\mbox{
\begin{picture}(200,100)(0,0)
\Text(-50,50)[]{(a)}
\Vertex(20,20){2}
\Vertex(20,80){2}
\Vertex(40,75){1}
\Vertex(43,72){1}
\Vertex(45,25){1}
\Vertex(47,27){1}
\Vertex(49,28){1}
\Vertex(70,70){1}
\Vertex(72,68){1}
\Vertex(100,47){1}
\Vertex(120,60){1}
\Vertex(130,30){1}
\Vertex(145,70){1}
\Vertex(170,25){2}
\Vertex(170,80){2}
\Vertex(105,85){1}
\Vertex(110,75){1}
\Vertex(130,95){1}
\DashLine(20,20)(20,80){4}
\DashLine(20,20)(40,75){2}
\DashLine(20,20)(43,72){2}
\Line(20,20)(45,25)
\Line(20,80)(40,75)
\Line(40,75)(43,72)
\DashLine(43,72)(45,25){2}
\Line(47,27)(45,25)
\DashLine(43,72)(47,27){2}
\DashLine(43,72)(49,28){2}
\Line(47,27)(49,28)
\Line(43,72)(70,70)
\Line(70,70)(72,68)
\DashLine(49,28)(70,70){2}
\Line(49,28)(100,47)
\DashLine(49,28)(72,68){2}
\DashLine(100,47)(72,68){2}
\Line(120,60)(72,68)
\Line(120,60)(130,30)
\DashLine(120,60)(100,47){2}
\Line(100,47)(130,30)
\Line(145,70)(170,80)
\Line(145,70)(170,25)
\DashLine(170,25)(170,80){4}
\DashLine(72,68)(105,85){2}
\Line(72,68)(110,75)
\DashLine(105,85)(110,75){2}
\Line(105,85)(130,95)
\Line(110,75)(130,95)
\Line(70,70)(105,85)
\LongArrow(0,50)(15,50)
\LongArrow(190,52)(175,52)
\Text(42,82)[]{$A$}
\Text(49,20)[]{$B$}
\Text(70,77)[]{$C$}
\Text(15,20)[]{$0$}
\Text(15,80)[]{$1$}
\end{picture}
}}
\scalebox{1.3}{\mbox{
\begin{picture}(200,100)(0,0)
\Text(-50,50)[]{(b)}
\Vertex(20,20){2}
\Vertex(20,80){2}
\Vertex(40,75){1}
\Vertex(43,72){1}
\Vertex(45,25){1}
\Vertex(47,27){1}
\Vertex(49,28){1}
\Vertex(70,70){1}
\Vertex(72,68){1}
\Vertex(100,47){1}
\Vertex(120,60){1}
\Vertex(130,30){1}
\Vertex(145,70){1}
\Vertex(170,25){2}
\Vertex(170,80){2}
\Vertex(105,85){1}
\Vertex(110,75){1}
\Vertex(130,95){1}
\DashLine(20,20)(20,80){4}
\DashLine(20,20)(40,75){2}
\DashLine(20,20)(43,72){2}
\Line(20,20)(45,25)
\Line(20,80)(40,75)
\Line(40,75)(43,72)
\DashLine(43,72)(45,25){2}
\Line(47,27)(45,25)
\DashLine(43,72)(47,27){2}
\DashLine(43,72)(49,28){2}
\Line(47,27)(49,28)
\Line(43,72)(70,70)
\DashLine(70,70)(72,68){2}
\DashLine(49,28)(70,70){2}
\Line(49,28)(100,47)
\DashLine(49,28)(72,68){2}
\DashLine(100,47)(72,68){2}
\Line(120,60)(72,68)
\Line(120,60)(145,70)
\DashLine(120,60)(130,30){2}
\DashLine(120,60)(100,47){2}
\Line(100,47)(130,30)
\DashLine(145,70)(170,80){2}
\DashLine(145,70)(170,25){2}
\DashLine(145,70)(130,30){2}
\DashLine(170,25)(170,80){4}
\Line(130,30)(170,25)
\Line(70,70)(105,85)
\DashLine(72,68)(105,85){2}
\Line(72,68)(110,75)
\DashLine(105,85)(110,75){2}
\Line(105,85)(130,95)
\DashLine(110,75)(130,95){2}
\Line(110,75)(145,70)
\DashLine(145,70)(130,95){2}
\Line(170,80)(130,95)
\LongArrow(0,50)(15,50)
\LongArrow(190,52)(175,52)
\DashLine(137,15)(137,98){8}
\Text(42,82)[]{$A$}
\Text(49,20)[]{$B$}
\Text(70,77)[]{$C$}
\Text(15,20)[]{$0$}
\Text(15,80)[]{$1$}
\end{picture}
}}
\caption{\label{figeffectivegluonsmult} (a) The emission of a new
  branch from a small dipole is suppressed, and proportional to the
  square of the small dipole size. However, if such a branch is
  emitted and interacts with a dipole from the target (b), then the
  small size has to correspond to a large $k_\perp$ in the final
  state.}}

\subsection{Gluon Recombination and Frame Dependence}
\label{sec:frame-dependence}

As mentioned in section \ref{sec:muell-dipole-form} the resulting
cross section is relatively insensitive to the reference frame in
which a collision is studied. If the interaction is studied in a frame
with rapidity $y$ relative to the projectile, then (in leading order)
the projectile cascade has evolved by a factor $e^{\lambda y}$ and the
cascade from the target by a factor $e^{\lambda (Y-y)}$, where $Y$
represents the total rapidity interval. The product is proportional to
$e^Y$, and thus independent of $y$. It is also demonstrated in
ref.~\cite{Mueller:1996te} that the result for a single chain is the
same in all frames, and independent of whether the cascades are
developed from the projectile end or from the target end. This is a
consequence of the M\"obius invariance of the process, and is exactly
true in the limit when the cutoff $\rho$ goes to zero.

However, including the unitarity corrections from multiple collisions
also implies that the result is no longer frame independent. The
contributions from multiple collisions in eq.~(\ref{multint}) are
formally colour suppressed $\sim 1/N_c^2$. What is treated as a
multiple collision in figure \ref{fig:multcoll} or
\ref{figeffectivegluonsmult} corresponds in the target rest frame to a
process where two dipoles fuse to a single dipole (or two gluons fuse
to a single gluon) before the collision with the target. Such
recombinations\footnote{In the terminology of
  ref.~\cite{Mueller:1996te} the effect of multiple collisions is
  called a unitarisation effect, while the effect of gluon
  recombination is called saturation. As the separation between the
  two mechanisms is not dynamical, but only a question of bookkeeping
  depending on the particular frame of reference used in the analysis,
  we do in this paper not make this distinction.} are consequently
also colour suppressed, and they are not included in the dipole
cascade evolution, which is only leading order in $N_c$. In the final
state this process gives a closed dipole loop, which is colour
disconnected from the rest of the system. In a string fragmentation
scenario it gives a closed string, which fragments as a separate
system.  This implies that such loops are only taken into account if
the collision is studied in a frame where they are appearing as
multiple collisions between two branches coming from each direction,
and not in a frame where they appear as gluon recombination, as e.g.\ 
in the target rest frame.

We conclude that within this formalism the unitarisation corrections
\textit{do} depend on the Lorentz frame used. As discussed in
ref.~\cite{Mueller:1996te}, for symmetric collisions the optimal frame
should be the overall rest system, where both the projectile and the
target may evolve, and the probability is largest that a dipole loop
corresponds to a multiple collision event.  This is illustrated in
figure \ref{fig:multcoll}, where in the overall rest system only one
loop does not correspond to a multiple collision but to a gluon
recombination. In a less central frame more loops would correspond to
recombinations and there would be correspondingly fewer multiple
sub-collisions.

The situation is different for onium scattering on a dense nuclear
target. Here the target is treated as a large number of dipoles, and
multiple collisions are most likely when different initial dipoles
from the target are involved. Therefore multiple collisions are well
accounted for in the target rest frame, where the projectile cascade
is fully developed. This is also the approach taken in the BK
equation, which similarly takes into account multi-pomeron exchange
but not the gluon recombination process representing pomeron fusion.

The frame independence is a very essential feature of the LDC model,
and we think it is important to develop a formalism in which multiple
collisions and gluon recombinations appear on an equal footing, in a
frame independent description. We will return to this problem in a
future publication.

There is also another frame dependent effect, which has a more
kinematic origin.  For a finite cutoff $\rho$, or for the effective
cutoff $\rho(\Delta y)$, the frame independence is only approximate,
also for a single chain. Furthermore, in our scheme for energy
conservation every new branch takes away energy, and therefore in a
cascade with many branches the energy in each individual branch is
reduced.  As discussed in section \ref{e-m-conservation}, a branch can
only be realized if the interaction with the target can provide the
necessary $p_-$ momentum. The other branches are virtual and cannot be
realized in the final state. This is e.g.\ the case for the branches
marked $B$ and $C$ in figure \ref{multcoll}.  As our constraint from
energy-momentum conservation also includes the fractions needed to
evolve the non-interacting branches the effect is somewhat
overestimated. Quantitatively this bias turns out to be small. For
dipole--dipole collisions, as described in section \ref{sec:res-oni-oni},
we find that the cross sections calculated in the cms at $y = 0.5 \,Y$
or asymetrically at $y = 0.75 \,Y$ differ by less than 4\%.

\section{The Monte Carlo Implementation}
\label{sec:monte-carlo-impl}

In this section we briefly describe the Monte Carlo scheme used to
calculate the results presented in this paper. As we have mentioned
before, the onium state is evolved in rapidity. For a given dipole one
then generates $y$ and $\mathbf{r}$ values for a possible gluon
emission (dipole splitting) using \eqref{splitprob}.

\subsection{Kinematics}

To be able to study the effects of energy-momentum conservation we
simply assign besides a transverse position and a rapidity, a positive
light-cone momentum and a transverse momentum to each parton in the
evolution, where $k_+=k_\perp e^{-y}=(2/r)e^{-y}$. The dynamical
cutoff is then given by $\rho =2e^{-\Delta y}/k_{\perp\mathrm{parent}}$. When a
dipole emits a gluon the mother partons will receive recoils from the
emitted gluon. Since a gluon belongs to two different dipoles one has
to decide how the emission of a gluon effects the neighboring dipoles.
We will simply assume that when a dipole emits a gluon the mother
gluons need to supply all the needed energy. Thus the next time a
neighboring dipole emits a gluon, the avaliable energy is reduced
because one of its gluons has lost energy from the earlier emission.

Consider the emission of gluon $n$ from the dipole $ij$ between
partons with light-cone momenta $k_{+i}$ and $k_{+j}$.
The transverse distances between the new gluon and
partons $i$ and $j$ are denoted $r_{in}$
and $r_{jn}$ respectively. We then assume that the nearest parent gluon 
takes the dominant fraction of the recoil. Thus if $k_{+n}$ is the 
momentum given to the emitted gluon, then the momenta left to the 
parents after the emission are given by
\begin{equation}
k_{+i}'=k_{+i}-\frac{r_{jn}}{r_{jn}+r_{in}}k_{+n} \; \; \textrm{and}\; \;
k_{+j}'=k_{+j}-\frac{r_{in}}{r_{jn}+r_{in}}k_{+n}.
\end{equation}
Alternative formulas for sharing the recoils have also been studied,
but the result does not depend sensitively on the exact formula. When
an emission is generated we always make sure that $k_+'\ge0$.

As we in this paper are not investigating final state properties but only total
cross sections, we will neglect the directions of the transverse momenta.
Keeping only the lengths of the $\mathbf{k}_\perp$ vectors, we neglect the 
possibility that two contributions may be of approximately equal size
in opposite directions, giving a much smaller vector sum. This approximation 
has to be improved in future analyses of final states, but should not be 
essential here. Thus in our approximation the transverse momentum of a 
parton will be
decided by the shortest distance to another parton, with which it has
formed a dipole, and when the gluon $n$ is emitted from the dipole 
$(ij)$, its transverse momentum is given by
\begin{equation}
k_{\perp n}=2\max\biggl(\frac{1}{r_{in}},\,\frac{1}{r_{jn}}\biggr).
\end{equation}
In analogy the recoils on the emitting partons are given by
\begin{eqnarray}
& &k'_{\perp i} =\max\biggl(k_{\perp i} ,\,\frac{2}{r_{in}}\biggr) \\
& &k'_{\perp j} =\max\biggl(k_{\perp j} ,\,\frac{2}{r_{jn}}\biggr). \nonumber
\end{eqnarray}

The recoil also implies that the rapidity is modified for the parents,
and is determined by the relation $y'=\log\frac{k_\perp'}{k_+'}$.
In some cases this could imply that an emitting parton ends up with
rapidity larger than the rapidity of the emitted gluon. Since the
cascade is assumed to be ordered in rapidity, such emissions are not
allowed, and we demand that $y_i',y_j'\leqslant y_n$. In this way we
also avoid the situation where there are partons which have rapidities
outside the allowed rapidity interval. As mentioned in section
\ref{e-m-conservation}, negative light-cone momentum is also
conserved.  This we do by imposing the condition
$k_{-n}\geqslant\max(k_{-i},k_{-j})$ during the evolution, where
$k_{-i}=2e^{y_i}/r_{ij}$ and $k_{-j}=2e^{y_j}/r_{ij}$ according to the
discussion in section \ref{sec:final-states}. For every generated
gluon one checks the kinematical constraints described above, and in
case one of them is not satisfied a new gluon is generated in a way
which automatically includes the same phase-space restrictions in the
integral of the Sudakov form factor.

The constraint on the negative light-cone momentum given above implies
that $k_-$ steadily increases. As also discussed in section
\ref{e-m-conservation} we presume that in the end the collision with
the target provides the necessary $k_-$ to put the dipole chain on
shell. To make sure that this indeed is possible, we impose the following
constraint on the colliding dipoles
\begin{equation}
\frac{16}{r_{ab}^2} < k_{+a} \cdot k_{-b}
\label{collision-constraint}
\end{equation}
Here $a$ and $b$ denote two colliding gluons, which are connected in
the recoupling as shown in figure~\ref{figgammagamma}. The left moving
onium is the one with larger $k_-$ while the right moving onium has
larger $k_+$.  When two dipoles collide there is only one possible way
to reconnect the gluons, which is consistent with the colour flow. The
constraint in eq.~(\ref{collision-constraint}) has to be satisfied for
both pairs of connected gluons.  If one of the constraints is not
satisfied, the corresponding scattering amplitude is set to zero,
which guarantees that only sub-collisions, which satisfy energy-momentum
conservation, contribute to the cross section.

As a final remark we mention that all calculations are performed using a
fixed coupling constant \as, corresponding to $\bar{\alpha}=0.2$. We intend
to study the effects of a running coupling in future investigations.

\section{Results}
\label{sec:results}

\subsection{Dipole Multiplicity}
\label{sec:multiplicity}

We will begin this section by describing some of the general
properties of the dipole evolution, and we first study how the dipole
multiplicity grows with energy. In figures \ref{nod} and \ref{pnod} we
show the average number of dipoles with and without energy
conservation. As some dipoles have to be regarded as virtual, according to 
the discussion in section \ref{sec:final-states}, these results do not have 
a direct physical interpretation. They are interesting because they may be
helpful, e.g. in estimates of effects of multiple collisions and the 
efficiency of the MC program.
Without energy conservation a fixed cutoff, $\rho$, is
needed for small dipole sizes, and in figure \ref{pnod} results are
shown for $\rho=0.04\, r_0$ and $\rho=0.02\, r_0$, where $r_0$ is the
size of the initial dipole starting the cascade. In all cases the
total dipole multiplicity is growing exponentially with rapidity.  A
small cutoff favors the production of very many small dipoles, which
is reflected in a very large dipole multiplicity, as seen in
figure \ref{pnod}.  With energy conservation the dynamical cutoff
$\rho(\Delta y)$, discussed in section~\ref{e-m-conservation}, is
large for small values of $\Delta y$, and this feature effectively
suppresses the production of many small dipoles in a small rapidity
interval. Comparing the two figures we see that energy conservation
indeed does have a very large effect. With energy conservation the
multiplicity at $Y\approx 10$ is a factor 20 below the result obtained
including energy conservation with the smaller cutoff value.

\FIGURE{
    \includegraphics[angle=270, scale=0.6]{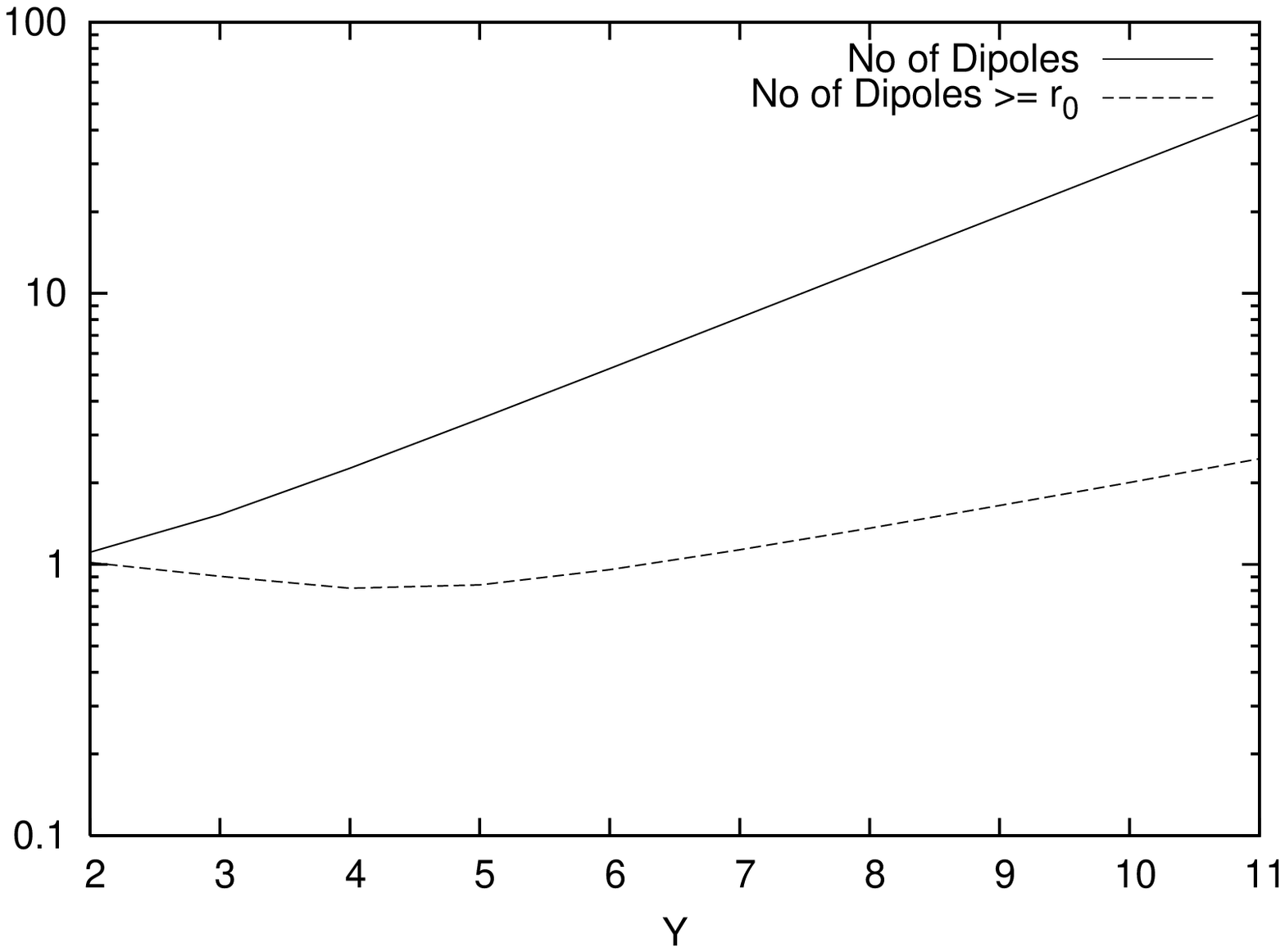}
    \caption{\label{nod}The average total number of dipoles (full line)
      together with the average number of large dipoles (dashed line)
      in the onium state when evolved with energy conservation.}  }

\FIGURE{
  \includegraphics[angle=270, scale=0.6]{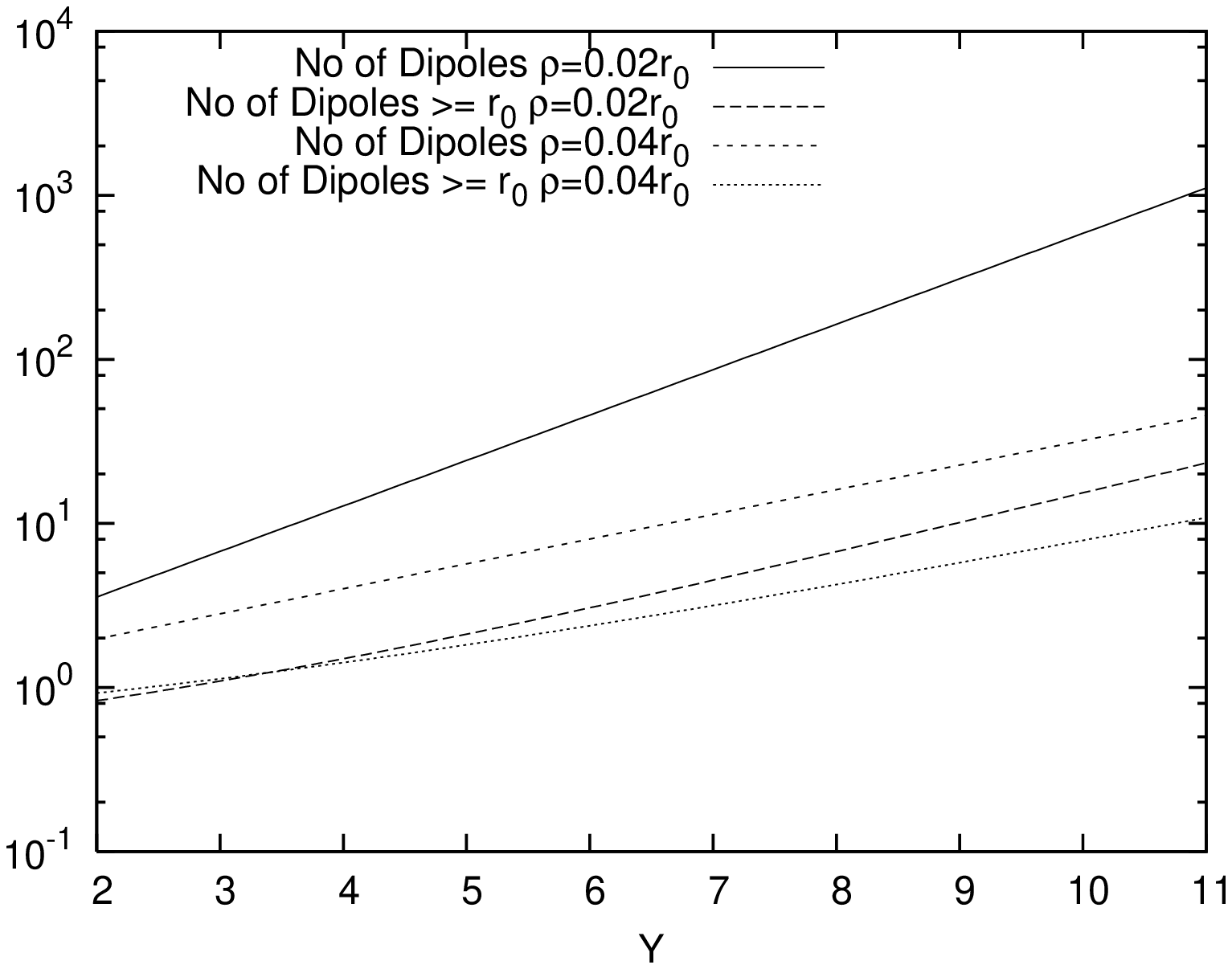}
  \caption{\label{pnod}The average total number of dipoles
    together with the average number of large dipoles in the onium
    state when evolved \emph{without} energy conservation and using two
    different cutoffs. For the smaller cutoff, $\rho=0.02$ the total
    number is given by the full line while the number of large dipoles
    is given by the dashed line. For the larger cutoff, $\rho=0.04$
    the total number is given by the short-dashed line while the
    number of large dipoles is given by the dotted line.}}

Without energy conservation the strong sensitivity to the small dipole
cutoff reflects the large probability to emit very small dipoles
(c.f.\ eq.~(\ref{splitprob})). As the small dipoles also have small
cross section, one could imagine that the differences seen in the
dipole multiplicity is rather unessential for total cross sections.
This is, however, not the case.  In figures \ref{nod} and \ref{pnod}
we also show the number of dipoles with sizes larger or equal to the
initial dipole size.  With energy conservation this number changes
rather slowly and exceeds one first at $Y \sim 7$, while without
energy conservation it is steadily increasing with energy. This is the
case also for the larger cutoff value, although in this case the total
multiplicity is not significantly larger than in the energy conserving
case. This feature is further illustrated in figure \ref{size}, which
shows the distribution in dipole size at $Y=6$ and $Y=8$.
Energy-momentum conservation does not only suppress small dipoles,
which we understand as a result of conservation of the positive
light-cone component, $p_+$, but there is also a suppression of large
dipoles, as a consequence of $p_-$-conservation.  Thus we conclude
that the implementation of energy conservation does not only have an
effect on very small dipoles, which turns out to be less unimportant
for the total cross sections, but indeed also has a very strong effect
on the main features of the evolution. This will be more clearly
illustrated in the following subsections.

\FIGURE{
  \includegraphics[angle=270, scale=0.6]{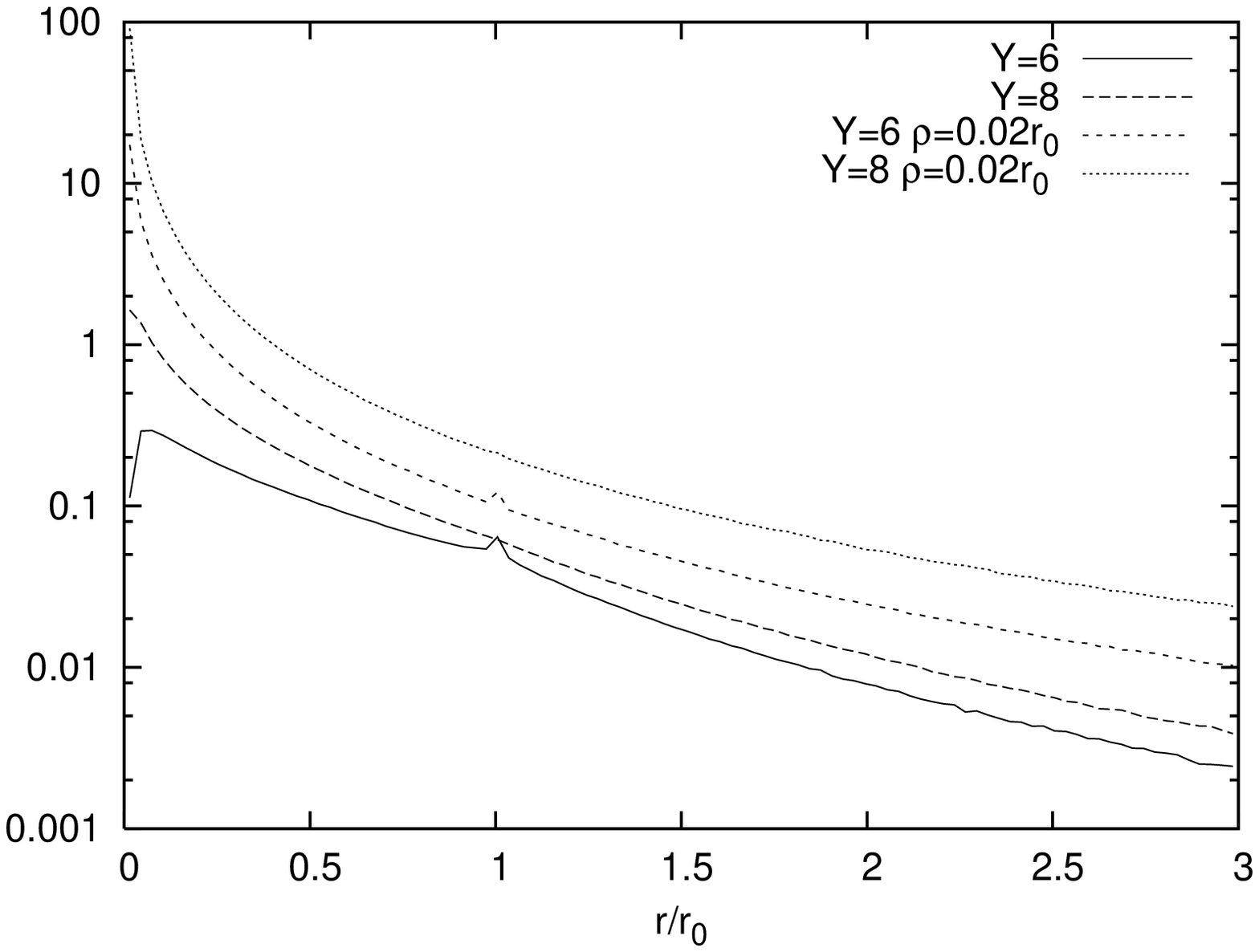}
  \caption{\label{size}The distribution in dipole size for $Y=6$ and 8.
    The solid and longdashed lines show the result from evolution with energy
    conservation while the shortdashed and dotted lines show the same for 
    evolution
    \emph{without} energy conservation with the cutoff $\rho = 0.02\,r_0$.}}

\subsection{Onium--Onium scattering.}
\label{sec:res-oni-oni}

We will here study the collision between two onium states, which we
regard as two incoming dipoles. We denote the initial dipole sizes
$r_1$ and $r_2$ respectively, and we imagine $r_2$ as the target
dipole with fixed size, while we vary the projectile size $r_1 \propto
1/\sqrt{Q^2}$. We note that with energy conservation and fixed $\as$
there is no external scale, and therefore the result for the scaled
cross section $\sigma/r_2^2$ does not depend on $r_1$ and $r_2$
separately, but only on their ratio.

\FIGURE{
  \includegraphics[angle=270, scale=0.6]{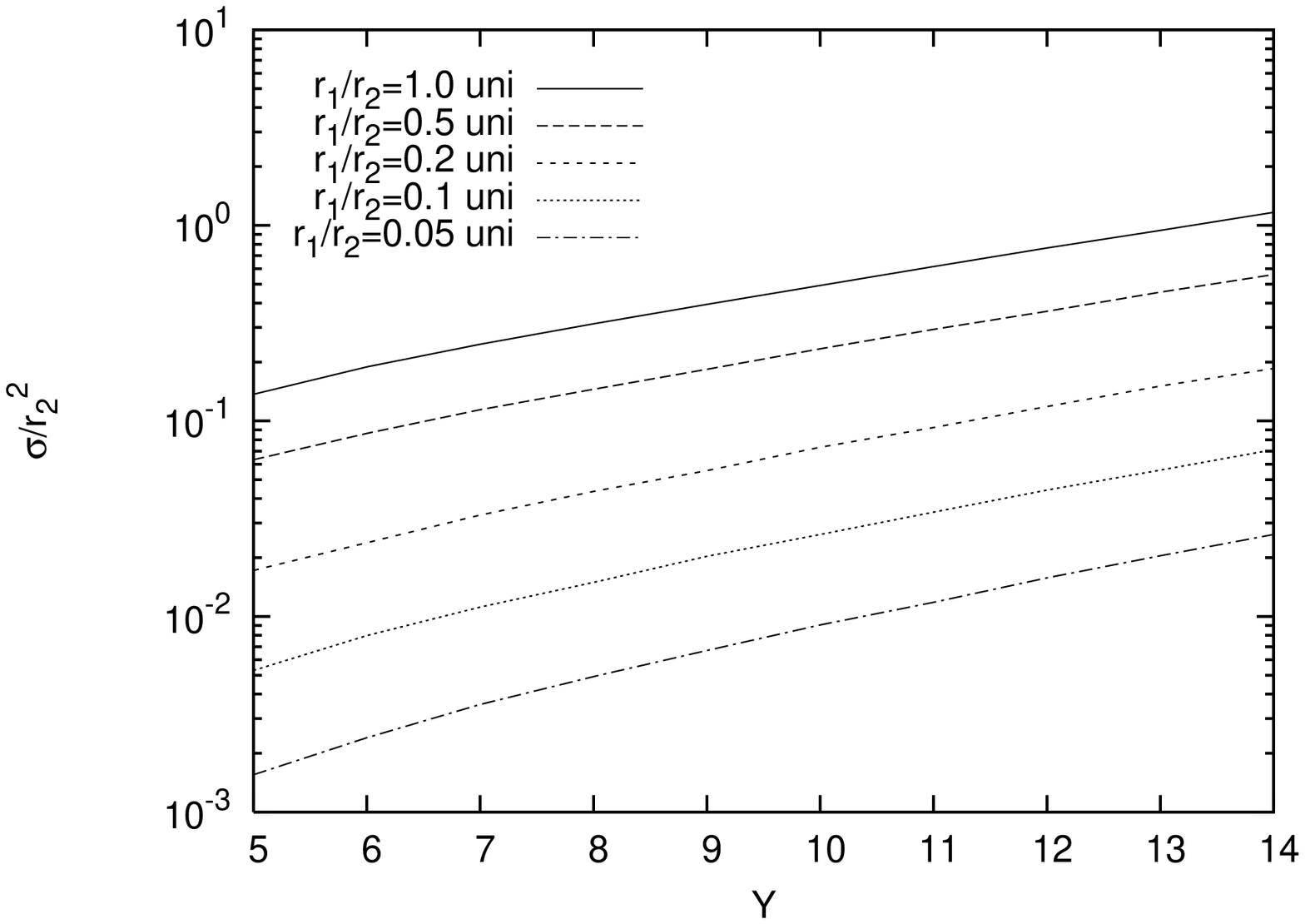}
  \caption{\label{ECuni}The scaled unitarised
    dipole--dipole cross section as a function of $Y$
    for different initial conditions.} }

\FIGURE{
  \includegraphics[angle=270, scale=0.6]{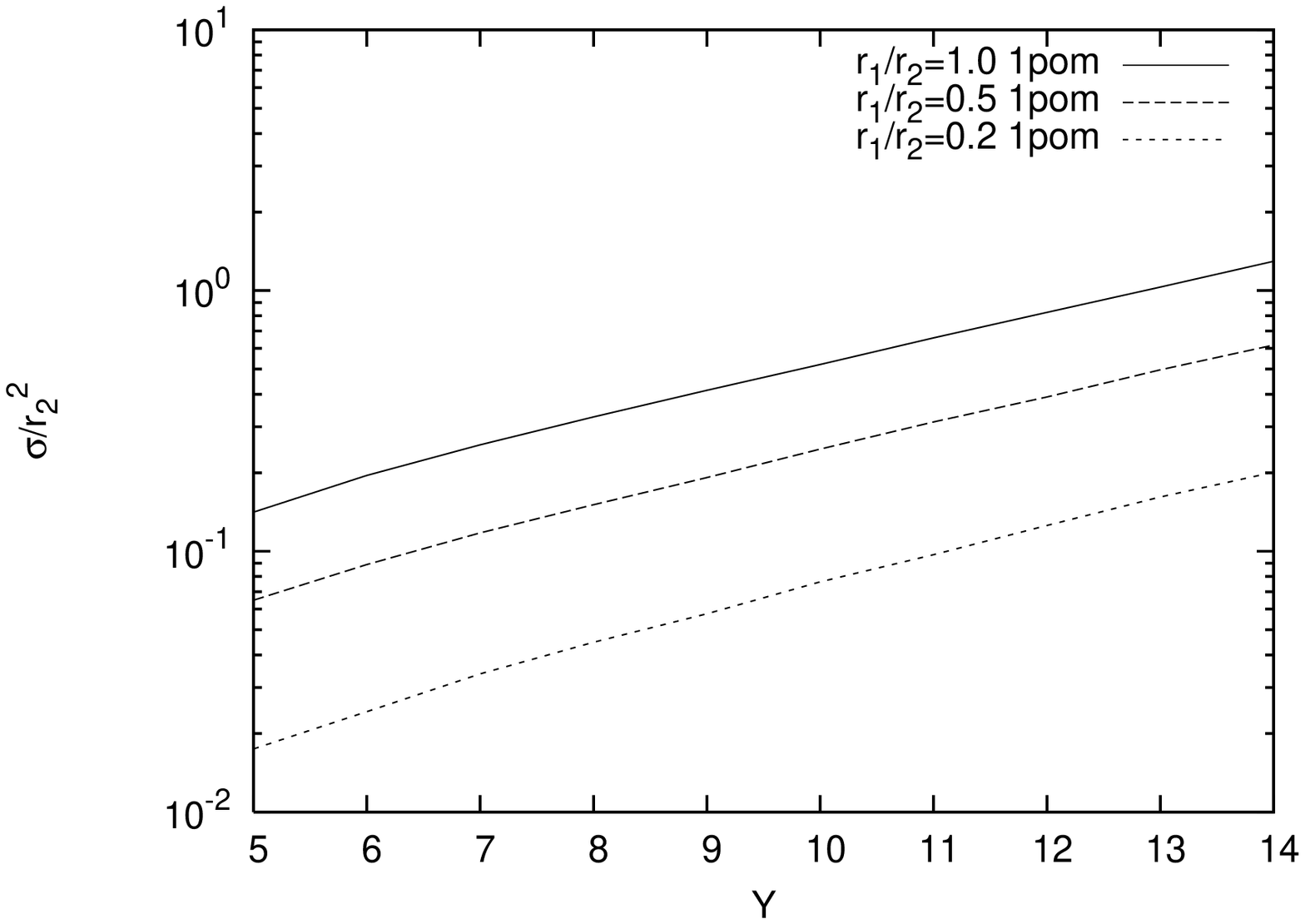}
  \caption{\label{EC1pom}The scaled one-pomeron
    dipole--dipole cross section as a function of $Y$
    for different initial conditions.} }
  
Figure \ref{ECuni} shows the total cross section as a function of the
rapidity $Y$ for different values of $r_1/r_2$, obtained including
energy conservation and unitarisation in accordance with
eq.~(\ref{multint}). The result from single pomeron exchange, where
the parenthesis in eq.~(\ref{multint}) is replaced by $\sum f_{ij}$,
is shown in figure \ref{EC1pom}, and we see that these results are
almost identical to those in figure \ref{ECuni}. We note in particular
that the curves are not straight lines, as is expected from leading
order BFKL. This implies that the effective slope,
$\lambda_{\mathrm{eff}}$, varies with rapidity, in a way expected as a
result of saturation. We also note that $\lambda_{\mathrm{eff}}$ grows
with larger values for the ratio between the dipole sizes.  This
effect is illustrated in figure \ref{lambdaoni}, and will be further
discussed in section~\ref{sec:res-ftwo-at-hera}.

\FIGURE{
  \includegraphics[angle=270, scale=0.6]{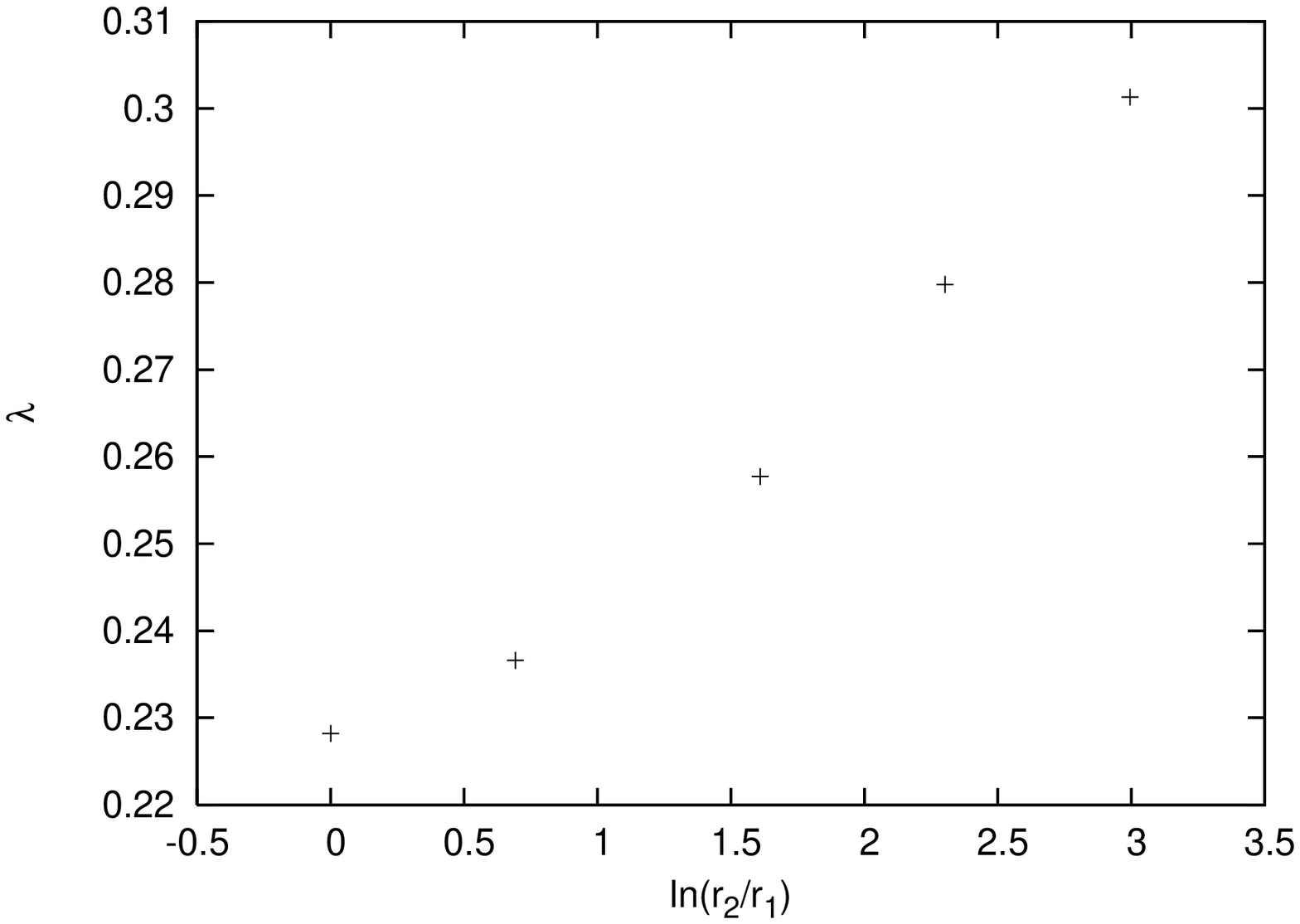}
  \caption{\label{lambdaoni}The effective power
    $\lambda_{\mathrm{eff}}$ calculated from the unitarised
    dipole--dipole cross section where energy conservation
    has been included.}  }

\FIGURE{
  \includegraphics[angle=270, scale=0.6]{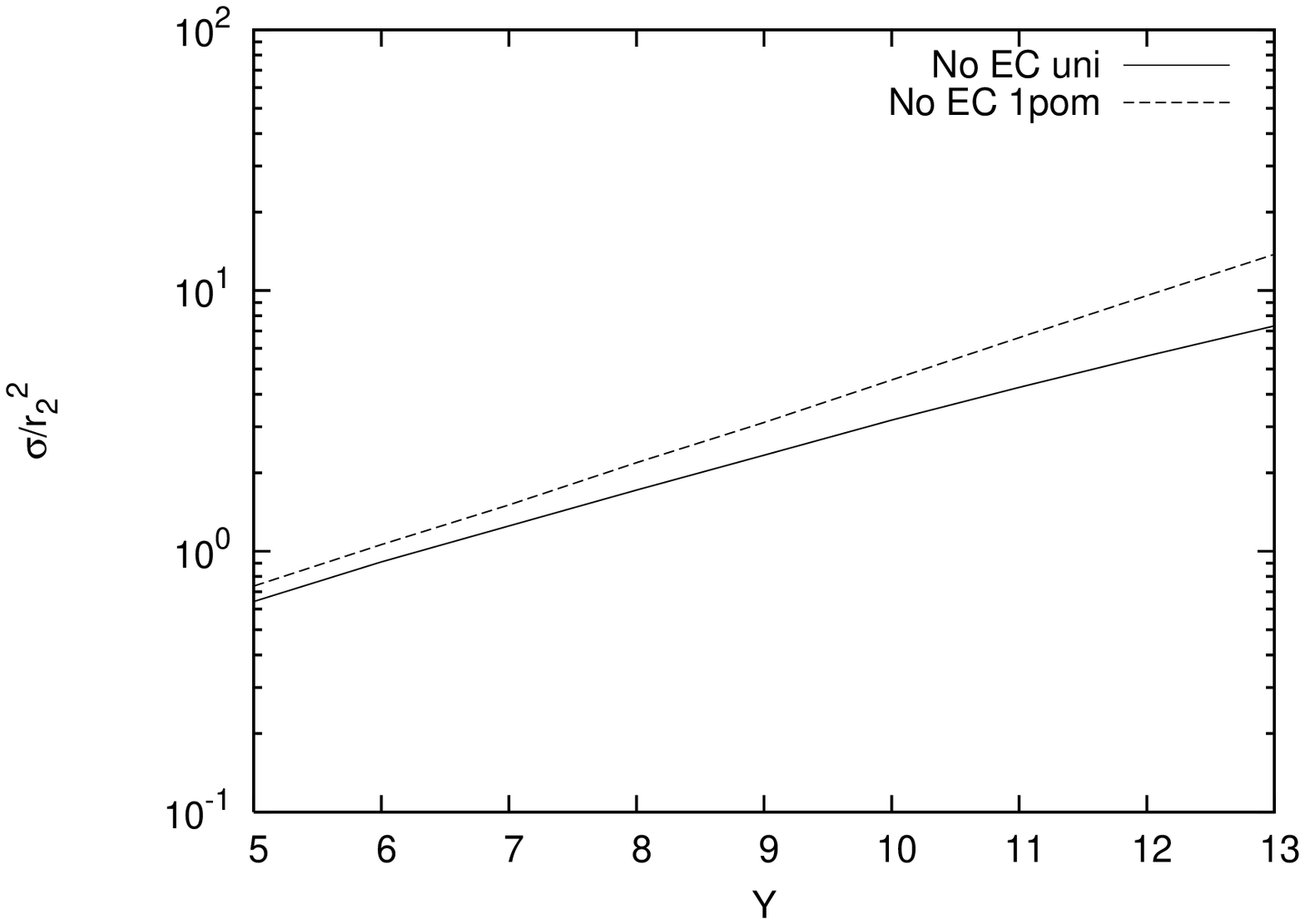}
  \caption{\label{noEC}The scaled unitarised (full line) and
    one-pomeron (dashed line) dipole--dipole cross
    sections calculated \emph{without} energy conservation.} }

For comparison, results obtained without energy conservation, with and
without unitarisation, are shown in figure \ref{noEC}. In this figure
the ratio $r_1/r_2$ is chosen equal to 1. We see that here the
one-pomeron cross section, without unitarisation, grows exponentially
with rapidity, proportional to $e^{\lambda Y}$ with a constant slope
$\lambda$.  Including unitarisation gives here a noticeable
suppression, which becomes stronger for larger rapidity and
correspondingly higher dipole density. This has the expected effect
that the growth rate is reduced for larger rapidities, with an
effective slope parameter $\lambda_{\mathrm{eff}}$ which is decreasing
for higher energies. Comparing figures \ref{ECuni} and \ref{EC1pom} we
note that already without unitarisation, the inclusion of energy conservation
also results in an effective slope, which is varying with energy in
much the same way.

\subsection{Dipole--nucleus scattering.}
\label{sec:res-oni-nuc}

Having studied dipole--dipole collisions we now apply our program to
dipole--nucleus collisions. We will focus on the qualitative features
and consider a toy model where the nucleus is given by a collection of
colour dipoles, which are distributed with a Gaussian distribution in
dipole size $\mathbf{r}$ and in impact parameter $\mathbf{b}$ and with
random relative angles. The number density of dipoles is given by
\begin{equation}
dN = B \cdot d^2\mathbf{r}e^{-\mathbf{r}^2/r_0^2}\cdot 
d^2\mathbf{b}e^{-\mathbf{b}^2/b_0^2}
\label{dN}
\end{equation}  
The parameters $r_0$ and $b_0$ are related to the estimated primordial
momentum in a proton and the nuclear radius respectively. As our model
is rather crude we have not tried to optimize these parameters, but
chosen the following canonical values: $r_0=1$~fm and
$b_0=A^{1/3}\cdot 1$~fm, $A$ being the mass number for the nucleus.
The normalization constant $B$ is determined by the requirement that
the transverse energy of the nucleus is set equal to $A\cdot 1$~GeV. To
simplify the calculations, the interaction amplitude for a
dipole--nucleus collision is calculated in the nucleus rest frame, by
convoluting the basic dipole--dipole amplitude with the distribution in
eq.~(\ref{dN}).  Although, as discussed in section
\ref{sec:frame-dependence}, the result is not exactly independent of
the Lorentz frame, the differences are not large, and should not be
essential for the qualitative studies in this section. For the
application to $ep$ scattering in the next section, where we will
compare our results with data from HERA, we will perform our
calculations in the hadronic rest system, which in that case should be
more accurate.

\FIGURE{
  \includegraphics[angle=270,scale=0.6]{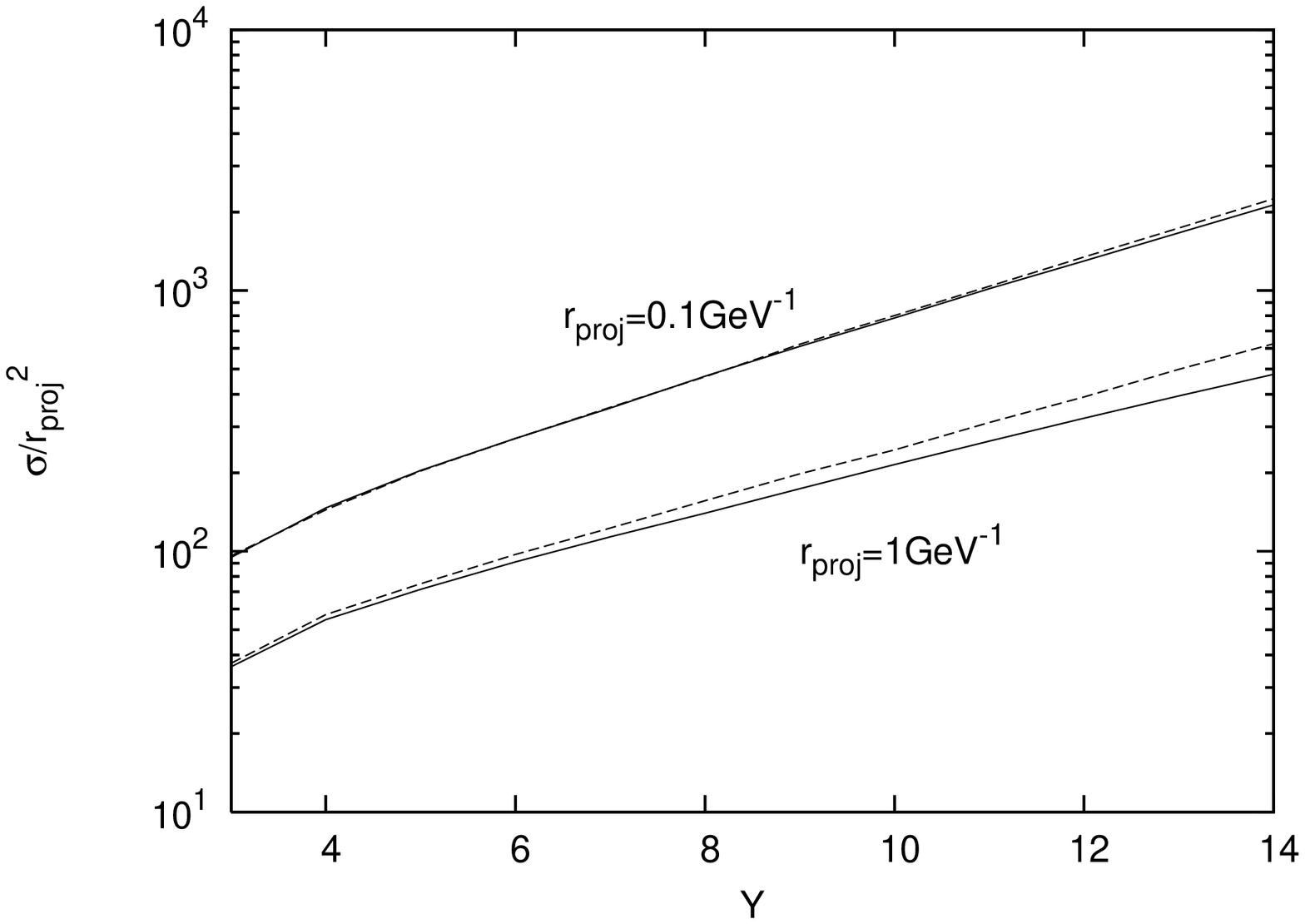}
  \caption{\label{sprobe}The dipole--nucleus cross section for 
    $r_{\mathrm{proj}}=0.1$ and $1$~GeV$^{-1}$ and $A = 200$. The
    unitarised result is shown by the solid lines, and the one-pomeron
    contribution by the dashed lines.}  }

The results for $A = 200$ and $r_{\mathrm{proj}}=0.1$ and
$1$~GeV$^{-1}$ are shown in figure \ref{sprobe}. Results are presented
both for single pomeron exchange and including unitarisation. The
effect of unitarisation grows with nuclear size and with the size of
the projectile.  For a small projectile of size $0.1$~GeV$^{-1}$ we
can see the effect of colour transparency, as the cross sections for
the unitarised and the one pomeron calculations are almost identical.
For a larger projectile we do see a clear effect from unitarisation,
but even for $r_{\mathrm{proj}}=1$~GeV$^{-1}$ and a nucleus with $A =
200$ this effect is only about $20$ percent in the rapidity interval
$10-14$. For smaller nuclei the effect will be correspondingly
smaller.

\FIGURE{
  \includegraphics[angle=270, scale=0.6]{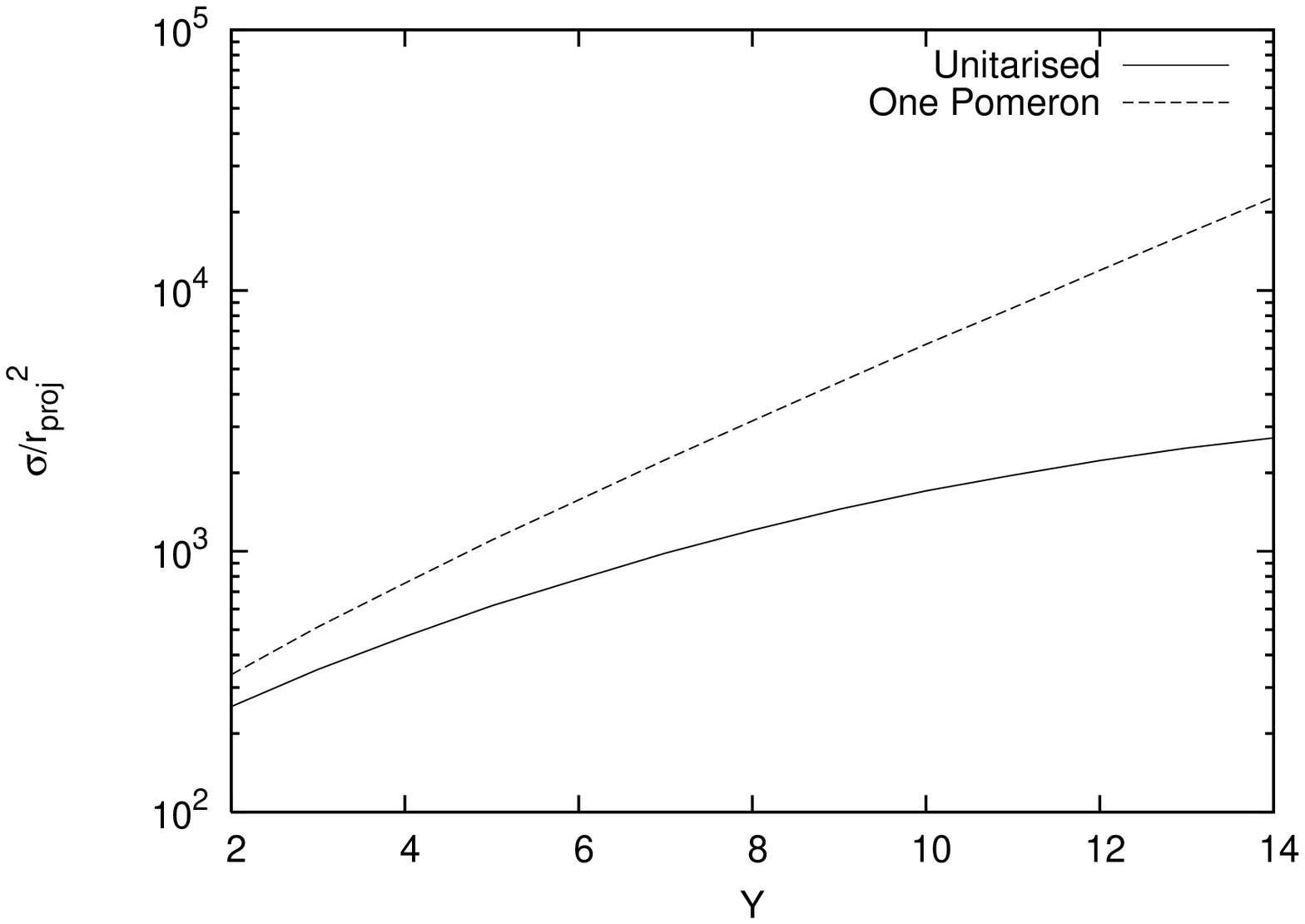}
  \caption{\label{pnucl}The scaled dipole--nucleus cross section,
    \emph{without} conservation of energy, for A = 200 with a projectile of
    size 1~GeV$^{-1}$.  The full and dashed line shows the result with
    and without unitarisation respectively.}  }

It is also interesting to study our toy model without energy
conservation, and figure \ref{pnucl} shows results for
$r_{\mathrm{proj}}=1$~GeV$^{-1}$ and $A = 200$, corresponding to the
larger projectile in figure \ref{sprobe}. The result is qualitatively
similar to the corresponding results for dipole--dipole collisions, in
the sense that the one-pomeron result is a straight line, while with
unitarisation the suppression is increasing for larger $Y$-values, and
the curve bends downwards. However, as expected the unitarisation
effect is here quantitatively much larger.

Comparing the results in figures \ref{sprobe} and \ref{pnucl} we see
that including energy conservation very strongly reduces the cross
section. This suppression becomes larger for higher energies, which
gives an effective slope, $\lambda_{\mathrm{eff}}$, which decreases
with energy in a way characteristic for saturation. The reduction of
the gluon density due to energy conservation is also so large that the
unitarity effects become comparatively small for available energies, even
for large nuclei.

\subsection[\ftwo at HERA]{\boldmath\ftwo at HERA}
\label{sec:res-ftwo-at-hera}

When we apply our model to deep inelastic ep scattering we want to
emphasize that we here only want to study the qualitative behavior.
We postpone a quantitative comparison with HERA data to a future
publication, where we can include effects of colour recombination and
improve the simple toy model for the proton target.

For the application to DIS ep collisions we here use the same toy
model described in section \ref{sec:res-oni-nuc}, with $A=1$. We also
identify $Q^2$ directly with $4/r\sub{proj}^2$ without taking into
account the detailed effects of the photon wavefunction.  This implies
that the number of dipoles in the target is much smaller than the
number of dipoles in an onium state developed to large $Y$-values as
described in section \ref{sec:multiplicity}. Hence, the collision is
more similar to the symmetric onium--onium scattering than to the very
unsymmetric onium--nucleus collision. To reduce the frame dependent
effects discussed in section \ref{sec:frame-dependence}, we therefore
study the dipole--proton collisions in the overall rest frame. We
neglect possible correlations between the target dipoles, which thus
are assumed to evolve independently. As the unitarisation effects turn
out to be small, and we here only study the total cross section, it is
also possible to neglect the fluctuations in the number of primary
target dipoles.

\FIGURE{
  \includegraphics[angle=270, scale=0.6]{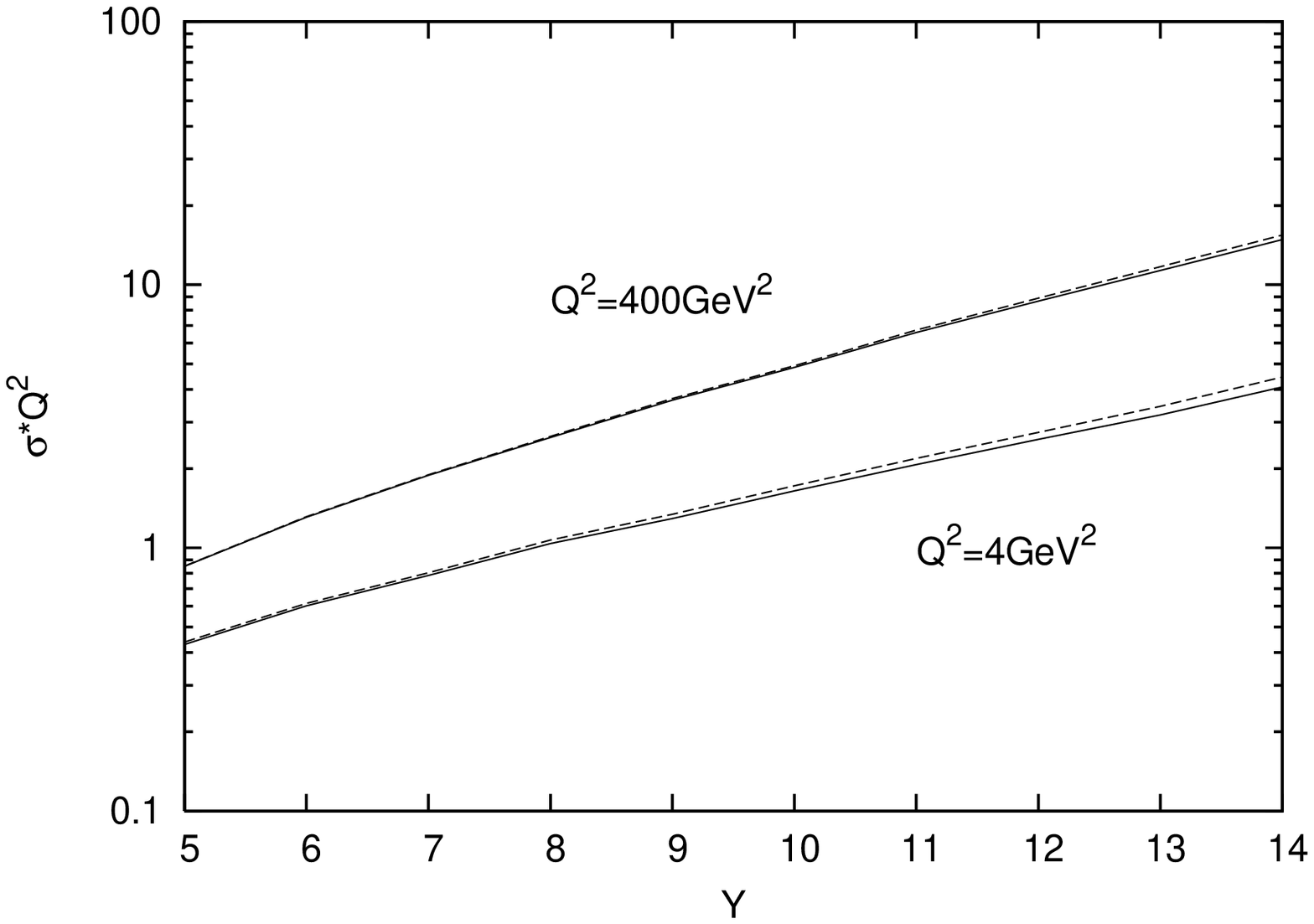}
  \caption{\label{labcm}The scaled dipole--p cross section as a
    function of log$1/x$, for $Q^2=4$~GeV$^2$ and $Q^2=400$~GeV$^2$.
    The unitarised results are shown by the solid lines while the
    dashed lines show the one-pomeron results.}  }

The resulting dipole--nucleon cross section is shown in
figure \ref{labcm} for two different projectile sizes, corresponding to
$Q^2=4\, \mathrm{GeV}^2$ and $Q^2=400 \,\mathrm{GeV}^2$. In this
figure we also show the result for single pomeron exchange, i.e.\ 
without unitarisation corrections, and we here see that the effect from
unitarisation is quite small. 

\FIGURE{
  \includegraphics[angle=270,scale=0.6]{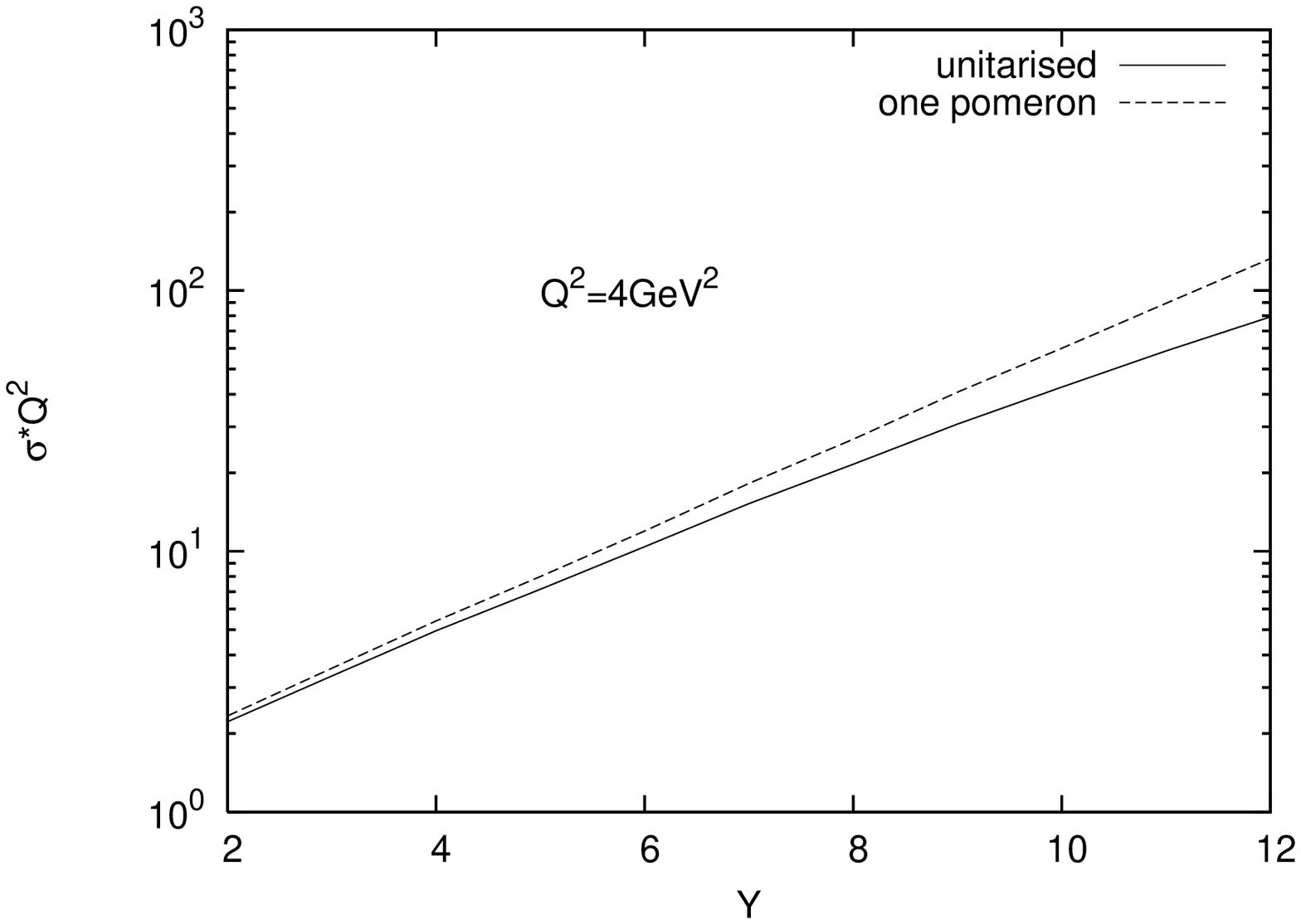} 
  \caption{\label{fig:sigma-ep-no-en-cons}The scaled dipole--p cross
    section as a function of log$1/x$ calculated \emph{without} energy
    conservation using $\rho=0.02$~GeV$^{-1}$. Both the unitarised
    (full line) and the one-pomeron (dashed line) calculations are
    shown.}  }

Fig.~\ref{fig:sigma-ep-no-en-cons} shows the corresponding results
without energy conservation. (The results presented here are obtained
for the cutoff $\rho=0.02~\mathrm{GeV}^{-1}$, and therefore somewhat
lower than the limiting values for $\rho \rightarrow 0$.)  We see that
without unitarisation and without energy conservation, the cross
section grows exponentially with $Y=\log 1/x$, or as a power of $x$.
With unitarisation (but without energy conservation) the growth rate
is, as expected, reduced and becomes continuously smaller with
decreasing $x$. We note, however, that energy conservation has a
similar effect, also without unitarisation, and the reduction in the
cross section due to energy conservation is so large that including
unitarisation does not have a significant effect.

In figure \ref{labcm} we also see that the logarithmic slope
$\lambda_{\mathrm{eff}}=d (\log \sigma)/d (\log 1/x)$ is increasing
with increasing $Q^2$.  As discussed above, $\lambda_{\mathrm{eff}}$
is not a constant for fixed $Q^2$, but depends on both $Q^2$ and $x$,
when unitarisation and/or energy conservation is taken into account.
To compare with experimental data we show in figure \ref{lambdaQ22}
$\lambda_{\mathrm{eff}}$ determined in the $x$-interval used in the
analysis by H1 \cite{Adloff:2001rw}, which varies from $x\approx
2\times 10^{-5}$ for $Q^2=1.5$~GeV$^2$ to $x\approx 3\times 10^{-2}$
for $Q^2=90$~GeV$^2$. We note that the result of our crude model
is not far from the experimental data, although the dependence on
$Q^2$ is somewhat weaker in the model calculations.  As in figure
\ref{labcm} we see that the effect of unitarisation is small, and, as
expected, it gets further reduced for larger $Q^2$-values. From figure
\ref{fig:sigma-ep-no-en-cons} we see that the result without energy
conservation and unitarisation corresponds to a much larger effective
slope, and also including unitarisation the result for
$\lambda_{\mathrm{eff}}$ is roughly a factor two larger than the
corresponding result in figure \ref{labcm}.

In conclusion we find that the result of our simple model is
surprisingly close to experimental data from HERA. This is very
encouraging, especially since we have not attempted to tune the model
in any way. The effect of energy conservation is a suppression for
small $x$-values and small $Q^2$, which is qualitatively similar to
the effect expected from unitarisation. This suppression is so strong
that the effect from adding unitarisation is only a very small
correction, visible for small  $Q^2$-values.
 
\FIGURE{
  \includegraphics[angle=270, scale=0.6]{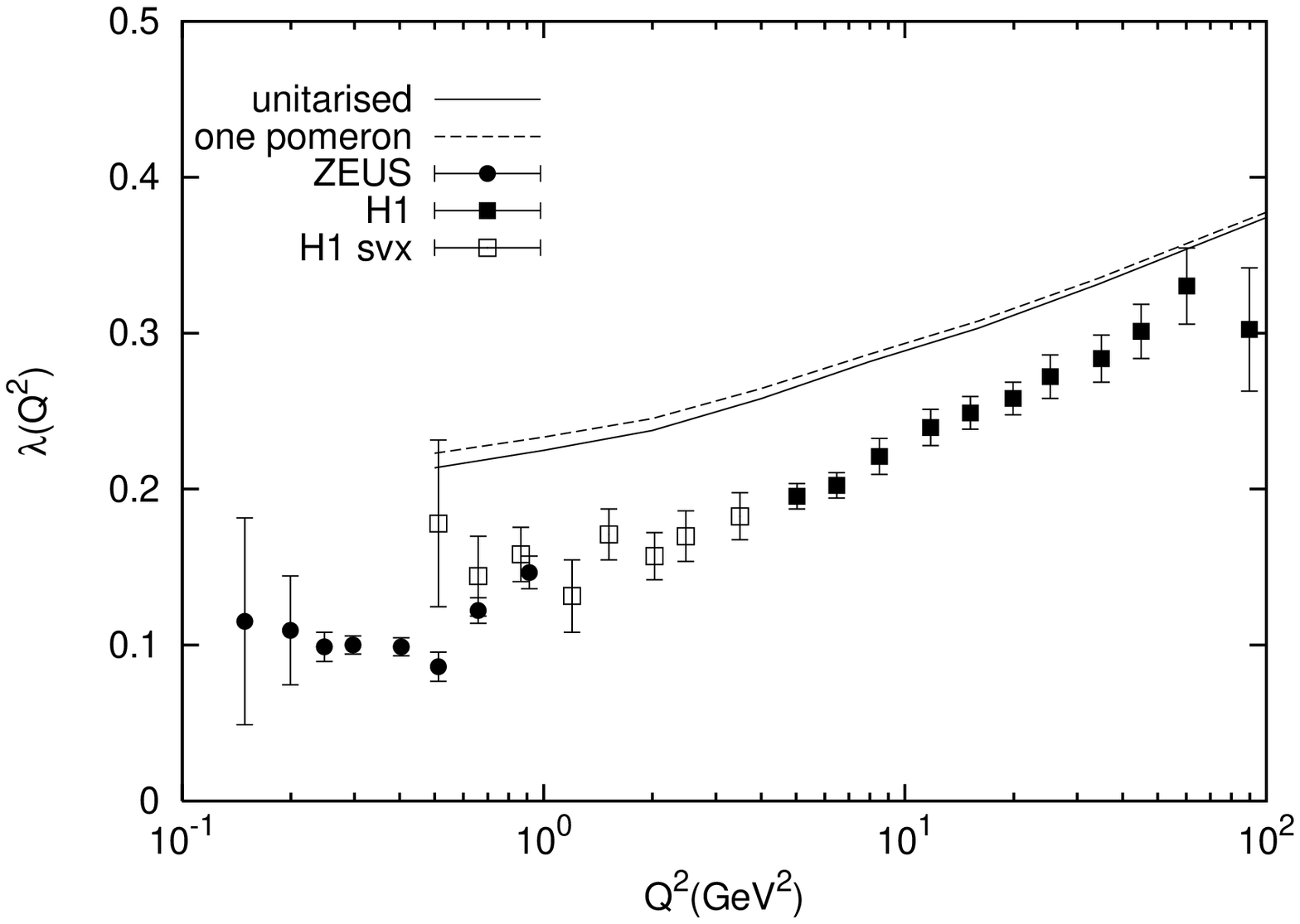}
  \caption{\label{lambdaQ22}The effective slope measured at different
    $Q^2$ compared to data from HERA. The full line is our model
    including unitarisation, while the dashed line is without. Filled
    circles are data from ZEUS\cite{Breitweg:2000yn},
    filled\cite{Adloff:2000qk} and open\cite{DIS04Petrukhin} squares
    are data from H1.}  }

\section{Conclusions}
\label{sec:conclusions}

Including both higher-order corrections and unitarisation effects in
the high-energy limit of QCD is not a simple task. Unitarisation
effects are more easily included in a dipole picture formulation in
transverse coordinate space, while higher-order corrections are more
easily formulated in transverse momentum space. In this report we have
used as a starting point that a large part of the NLO corrections to
BFKL are due to effects of energy-momentum conservation, which again
are more easily formulated in transverse momentum space. However,
after noting similarities between the LDC and Mueller dipole
formulations of high energy QCD, we conjecture that also in the latter
case, the essential contribution to the cross section comes from a
subset of dipoles between real final-state gluons, which necessarily
must respect energy and momentum conservation.

We have presented a way to implement energy and momentum conservation
in the Mueller dipole model. This is a necessary, although not
sufficient, requirement for selecting only final-state gluons, and
should also include most NLO corrections to the BFKL evolution.
Our way to implement energy-momentum conservation also eliminates the need for
a cutoff for small dipoles in Mueller's formalism, in which the large
number of small dipoles causes problems for a numerical treatment. 
Thus the number of dipoles produced in the evolution is drastically
reduced. This applies not only to the number of very small dipoles,
which do not much affect the resulting cross sections. Also the
number of large dipoles is reduced, resulting in a drastic reduction of the
cross section and in the effective slope $\lambda_{\mathrm{eff}} =
d\log{F_2}/d\log{x}$.

Also in standard BFKL evolution one would expect a large reduction of
the effective slope due to unitarisation effects. In our case the
growth of the cross section is already damped, making the inclusion of
unitarisation a rather small effect for the total cross section, even for deeply inelastic virtual
photon scattering on large nuclei. 

Comparing with the results of \cite{Gotsman:2005vc}, we find a larger
effect from energy-momentum conservation. One reason seems to be the
inclusion of $p_-$-conservation, which in our formalism is found to
have an important effect. Thus we find that including only
conservation of $p_+$, and not of $p_-$, increases the cross section
by a factor 2 (3) for dipole--proton collisions at $Q^2=4$ (400)GeV$^2$. 
Conservation of $p_-$ is related to the so
called consistency constraint\cite{Kwiecinski:1996td}, and in
\cite{Gotsman:2005vc} this contribution to the NLO BFKL kernel is
neglected, with the motivation that saturation effects suppress the
gluon density so that this contribution is less important. Naturally,
what is physically relevant is only the combined effect of both energy
conservation and saturation. Including the contributions in different
order, will also give different weights to the two effects. We feel
that energy and momentum conservation is the more fundamental
phenomenon, and hence obtain a much smaller effect from saturation.

To investigate the scattering on nuclei we have introduced a toy
model, where the nucleons are treated as a collection of dipoles with
a Gaussian distribution in sizes and impact parameter. We also use
this to model the scattering on a single nucleon and compared our
results with HERA data. Although we made no tuning of the parameters
of our model, we obtain a good semi-quantitative description of
the effective slope, $\lambda_{\mathrm{eff}}$, measured at low $x$ and
$Q^2$ at HERA.

Thus encouraged we will now continue to develop our model, and there
are several things which we would like to improve. A major development
would be to achieve a formalism which is completely frame independent.
This would entail the inclusion of true gluon recombinations in the
dipole evolution, and also a better understanding of the ``effective''
gluons with small transverse separation described in section
\ref{sec:final-states}. In this way we hope to also be able to use our
formalism to describe exclusive final-state properties. Another
important development would be to include effects of a running \as,
and also to improve our nucleon toy model to investigate the impact of
our formalism in nucleus--nucleus collisions. We intend to return to
all these issues in forthcoming publications.

\done{Large NLO corrections to BFKL, mostly due to energy conservation.}

\done{Standard dipole formalism is equivalent to LO BFKL, and gives a
  too rapid rise of the gluon density.}

\done{The large density gives rise to saturation effects.}

\done{Can saturation be seen at HERA?}

\done{Including energy conservation slows down rise of gluon density.}

\done{No need for saturation at HERA}

\done{Energy conservation is overestimated since virtual dipoles
  should not be affected. We return to this in future paper.}

\done{Gluon recombination is not taken into account. We return to this
  in future paper.}

\done{Running \as is not taken into account. We return to this
  in future paper.}

\bibliographystyle{utcaps}  
\bibliography{references,refs} 

\end{document}